\documentclass[aps,prd,preprint,groupedaddress]{revtex4-1}
\usepackage[dvipdfmx]{graphicx,color}
\usepackage{amsmath,amssymb}
\usepackage{bm}

\begin{document}

\preprint{KEK-TH-1914 and J-PARK-TH-0056}

\title{Unravelling the physical meaning of the Jaffe-Manohar
decomposition of the nucleon spin}


\author{M.~Wakamatsu}
\email[]{wakamatu@post.kek.jp}
\affiliation{KEK Theory Center, Institute of Particle and Nuclear Studies, High Energy Accelerator
Research Organization (KEK), 1-1, Oho, Tsukuba, Ibaraki 305-0801, Japan}



\begin{abstract}
A general consensus now is that there are two physically inequivalent
complete decompositions of the nucleon spin, i.e. the decomposition of
the canonical type and that of mechanical type. 
The well-known Jaffe-Manohar decomposition is of the
former type. Unfortunately, there is a wide-spread misbelief that this
decomposition matches the partonic picture, which states that
motion of quarks in the nucleon is approximately free. 
In the present monograph, we reveal that
this understanding is not necessarily correct and that the Jaffe-Manohar
decomposition is not such a decomposition, which natively reflects
the intrinsic (or static) orbital angular momentum structure of the nucleon.

\end{abstract}

\pacs{11.15.-q,12.38.-t, 12.20.-m, 14.20.Dh}

\maketitle


%
\section{Introduction}
\label{intro}

Over the past few years, there have been intensive debates on
the question whether the gauge-invariant complete decomposition
of the nucleon spin is possible or not.  
(See \cite{Review_LL14,Review_Waka14,Review_LL16}, for review.)
One of the central issues of this debate was concerned with the
significance of the concept of physical component of the gauge
field, which was first introduced by Chen et al. into the nucleon
spin decomposition problem \cite{Chen08,Chen09}. A consensus now is that the
definition of the physical component of the gauge field $A^\mu_{phys}$
is not unique. The ultimate reason is because $A^\mu_{phys}$ cannot
be defined independently of the choice of the Lorentz frame \cite{Waka15}.
The original proposal for $A^\mu_{phys}$ by Chen et al. amounts to
a nonabelian generalization of the familiar transverse-longitudinal decomposition
of the photon field also called the Helmholtz decomposition.
This Helmholtz decomposition works perfectly in the decomposition
problem of the total photon angular momentum into its intrinsic spin
and orbital parts \cite{VanEN94A,VanEN94B,BA1994,BAOA10}. The reason for it is twofold.
First, in this problem, we are dealing with free photons, or more
precisely, a wave packet of free photons. Second, the measurement of the
photon spin and orbital angular momentum (OAM) is carried out
in a fixed or prescribed Lorentz frame by making use of interactions with atoms, 
so that a particular choice of a Lorentz frame is nothing
problematical \cite{BBN13}. 

Unfortunately but importantly, the
situation is fairly different for the nucleon spin decomposition
problem. Here, we must handle quarks and gluons tightly bound
in the nucleon. To our present knowledge, the only one way
to probe the internal spin and OAM contents of such a composite
particle is to use deep-inelastic-scatterings (DIS).
One important property of DIS observables (or quasi-observables)
typified by parton distribution functions is the Lorentz-boost invariance
along the direction of the momentum of the parent nucleon \cite{Book_Collins11}.
Accordingly, the definition of $A^\mu_{phys}$, which is relevant for
the DIS measurements of the nucleon spin contents, must also
have this property \cite{Waka15}. (This is clear, for example, from the fact that
the measurable gluon spin is the first moment of the longitudinally polarized
gluon distribution function.)
The Coulomb-gauge-motivated definition of $A^\mu_{phys}$
proposed by Chen et al. does not satisfy this property \cite{Chen08,Chen09}.
The definition of $A^\mu_{phys}$, which satisfies this property, is
the light-cone-gauge motivated definition proposed by Hatta \cite{Hatta11}.
In this way, the claim that there can be infinitely many gauge-invariant
decompositions of the nucleon spin loses its basis, once the importance of
the boost-invariance requirement mentioned above is properly
recognized \cite{Waka15}.

Still, we are left with physically inequivalent two types of complete
decompositions of the nucleon spin, which are now called the canonical-type
decomposition and the mechanical-type (or kinetic-type) decomposition.
Note that, from the physical viewpoint, the canonical
decomposition is nothing different from the famous Jaffe-Manohar
decomposition \cite{JM1990} later refined by Bashinsky and Jaffe \cite{BJ1999}. 
In a series of paper \cite{Waka10,Waka11A,Waka12,Review_Waka14,Waka11C}, we
have advocated a view which favors the mechanical decomposition
rather than the canonical one as a natural decomposition
of the nucleon spin. 
Unfortunately, there still remains a wide-spread misbelief in the DIS
community that, as compared with the mechanical decomposition,
the Jaffe-Manohar decomposition is more compatible with the
familiaqr partonic picture of the quark motion inside the
nucleon \cite{JZZ13,Leader11, BJ1999}.
Undoubtedly, this misbelief comes from a careless extension of the parton
model idea, which states that the motion of partons in the nucleon is
free at the leading-twist approximation. Here is a pitfall, however. 
The partonic picture is certainly established for the collinear motion of
constituents along the direction of the nucleon momentum. 
As a matter of course, however, the generation of the OAM component
along the nucleon momentum requires motion of partons in the plane
perpendicular to this direction. Whether this transverse motion of
quarks is also partonic or not is a highly nontrivial question, which must
be judged only after careful consideration.
In fact, more natural picture is that this motion of quarks,
which generates the longitudinal-component of the OAM, is a circular
motion in the transverse plane. It seems obvious that
such a circular motion cannot be a free motion in any sense.

Anyhow, the above consideration throws a strong doubt on the
partonic interpretation of the Jaffe-Manohar decomposition of
the nucleon spin. What is a correct physical interpretation of
the quark and gluon OAM terms appearing in the Jaffe-Manohar
decomposition, then ? Is it really an observable decomposition ?
The purpose of the present paper is to answer these questions
as clearly as possible.
To this end, we think it very important to clearly understand
the distinction between the canonical OAM and the mechanical OAM
under the presence of the electromagnetic potential.
The famous Landau problem is a quantum mechanics of a
charged particle motion under the presence of uniform
magnetic field \cite{Landau1930}.
In sect.II, we concisely review the essence of this topics with a particular
intension of unmasking the identities of the two types of OAMs, i.e.
the canonical and mechanical OAMs.     
Next, in sect.III, we demonstrate an important role of the non-abelian
Stokes theorem in the nucleon spin decomposition problem following
the recent suggestion by Tiwari \cite{Tiwari15}. 
We explicitly show that the relation
between the canonical and mechanical OAMs derived by Burkardt
can more quickly be obtained by making use of this general
theorem \cite{Burkardt13}.
Next, after these preparations, we revisit in sect.IV several fundamental
questions of the gauge-invariant nucleon spin decomposition problem.
Can one say that the complete decomposition of the nucleon spin based
on the concept of the physical component of the gauge field is genuinely
gauge-invariant ? Which of the canonical type or the mechanical
type can be thought of as an observable decomposition ?
Next, in sect. V, we reveal the physical meaning of the
Jaffe-Manohar decomposition in a coherent fashion, to show
why its partonic interpretation is not justified.
Finally, in sect.VI, we summarize what we have clarified in the present
paper.

\section{Two orbital angular momenta in the Landau problem}
\label{sec:2}

In several previous publications, we repeatedly emphasized the fact that, under
the presence of strong background of magnetic field, what describes the
physical orbital motion of a charged particle is the mechanical (or kinetic)
OAM not the canonical one \cite{Review_Waka14,Waka10,Waka11A}. 
In view of the existence of strong color magnetic
field inside the nucleon as a quark-gluon composite,
this naturally implies that the physically favorable decomposition of the nucleon
spin is the mechanical (or kinetic) type decomposition not the canonical type one.  
Unfortunately, this reasonable claim is not necessarily accepted by many
of the deep-inelastic-scattering (DID) physics community.
This is due to a blind belief of the parton picture, which states that the motion
of quarks inside the nucleon must be approximately free at the leading order.
To correct this misunderstanding, we think it useful to understand
the essence of the famous Landau problem \cite{Landau1930}, i.e. the motion of
a charged particle in uniform magnetic field, especially by paying attention
to the physical content of the two OAMs. i.e. the canonical and 
mechanical OAMs \cite{Landau_Lifschitz,LW1999,FL00}.
(Very comprehensible lecture note on Landau problem can be found in
\cite{Murayama}.)
 
For simplicity, let us confine ourselves to the 2-dimensional motion of a particle
with charge $e$ in the $x$-$y$ plane under uniform magnetic field 
$\bm{B} = B \,\bm{e}_z$ along the $z$-axis. (Here, for clarity, the charge $e$ of the
particle is assumed to be positive.)
In classical mechanics, the Lorentz force causes a circular motion of the charged
particle.  The balance equation between the centrifugal force and the Lorentz force
reads as 
\begin{equation}
 \frac{m \,v^2}{r} \ = \ e \,B \,v .
\end{equation}
(Here and hereafter, we use the natural unit $c = \hbar = 1$.) 
This gives the radius of the circular motion : 
\begin{equation}
 r \ = \ \frac{m \,v}{e \,B},
\end{equation}
which is called the Larmor radius or the cyclotron radius. The energy of the system is given by
\begin{equation}
 E \ = \ \frac{1}{2} \,m \,v^2 \ = \ \frac{1}{2} \,m \,r^2 \,\omega^2,
\end{equation}
with $\omega = \frac{v}{r} = \frac{e \,B}{m}$ being the angular frequency of the cyclotron
motion. In classical mechanics, the cyclotron radius as well as the velocity $v$ can take
any real values. In quantum mechanics, the orbit of the cyclotron
motion  as well as the energy are quantized.
It can be seen already in the semi-classical treatment, which
corresponds to imposing the so-called Bohr-Sommerfeld quantization condition as
\begin{equation}
 \frac{1}{2 \,\pi} \,\oint \,\bm{p} \cdot d \bm{r} \ = \ n \ + \ \frac{1}{2} \, .
\end{equation}
With use of the relation $\bm{p} = m \,\bm{v} + e \,\bm{A}$, where $\bm{A}$ is the gauge
potential corresponding to the magnetic field $\bm{B}$, this leads to
\begin{eqnarray}
 n \ + \ \frac{1}{2} \ &=& \ \frac{1}{2 \,\pi} \,\oint \,(\,m \,\bm{v} \ + \ e \,\bm{A} \,) \cdot
 d \bm{r} \nonumber \\
 \ &=& \ m \,v \,r \ - \ \frac{1}{2 \,\pi} \,e \,B \,( \pi \,r^2) 
 \ = \ m \,v \,r \ - \ \frac{1}{2} \,m \,v \,r \ = \ \frac{1}{2} \,m \,v \,r , \label{Eq:BS_integral}
\end{eqnarray}
where we have used the relation $\omega = \frac{v}{r} = \frac{e \,B}{m}$.
Here use has been made of the Stokes theorem.
(The origin of the minus sign in front of the second term on the r.h.s. of Eq.(\ref{Eq:BS_integral})
is that the cyclotron motion is clockwise for $e \,B > 0$, and in this case the line integral of the
vector potential gives the negative of the magnetic flux inside the Larmor radius.) 
Multiplying both sides with $\omega$, this gives the quantized energy as
\begin{equation}
 \left( n \ + \ \frac{1}{2} \right) \,\omega \ = \ \frac{1}{2} \,m \,v^2 \ = \ E ,
\end{equation}
with $n$ being a non-negative integer. 
Accordingly, the cyclotron radius is also quantized as
\begin{equation}
 r_n \ = \ \sqrt{\frac{1}{e \,B}} \cdot \sqrt{2 \,n + 1} .
\end{equation}
As we shall see shortly, in quantum mechanics, above discrete orbit with the radius
$r_n$ corresponds to a Landau state describing quantized cyclotron motion.
However, we shall also see that each state has an infinite degeneracy originating from
the fact that each state with a definite Landau quantum number $n$ contains infinitely
many states, which are characterized by another integer $m$, the eigenvalue of
the canonical angular momentum operator $L_{can}$.  

As is well-known, a quantum mechanical treatment of the cyclotron motion requires to
introduce the vector potential $\bm{A}$, which is defined through the
relation $\nabla \times \bm{A} = \bm{B}$. 
The relevant Hamiltonian of the Landau problem is then given by
\begin{equation}
 H \ = \ \frac{1}{2 \,m} \,\bm{\Pi}^2 \ = \ \frac{1}{2 \,m} \,
 (\bm{p} \ - \ e \,\bm{A})^2 . \label{Eq:Landau_Hamiltonian}
\end{equation}
The choice of $\bm{A}$, which gives the same $\bm{B}$, is not unique but the
physics must be independent of this choice.
We say that the theory has a gauge invariance.
To solve the quantum mechanical problem explicitly, however, we are forced to
take some specific choice for the vector potential $\bm{A}$, which amounts to
taking a particular gauge choice. Some of the popular choices are
the rotationally-symmetric gauge given by
\begin{equation}
 \bm{A} \ = \ (A_x, A_y) \ = \ \frac{B}{2} \,(- \,y, x) , 
\end{equation}
the gauge with the translational-invariance along the $x$-axis given as
\begin{equation}
 \bm{A} \ = \ (A_x, A_y) \ = \ B \,(- \,y, 0) , \label{Eq:Landau_gauge}
\end{equation}
and the gauge with the translational-invariance along the $y$-axis given as
\begin{equation}
 \bm{A} \ = \ (A_x, A_y) \ = \ B \,(0, x) .
\end{equation}
The choice (\ref{Eq:Landau_gauge}) is the gauge used by Landau in solving
the problem for the first time, so that we call it the Landau gauge
hereafter \cite{Landau1930}.
Although gauge-invariant quantities are independent of the gauge choice,
the symmetric gauge is most convenient for understanding the relation between
the canonical OAM and the mechanical OAM, so that let us first 
work in this gauge. 

In the combination
\begin{equation}
 \bm{\Pi} \ = \ \bm{p} \ - \ e \,\bm{A} ,
\end{equation}
which enters the Hamiltonian, $\bm{p}$ is the standard canonical momentum,
while $\bm{\Pi}$ is called the mechanical (or kinetic) momentum. 
At variance with the canonical momenta, the mechanical momenta do not
commute with each other. Their commutation relation is
\begin{equation}
 [\Pi_x, \Pi_y] \ = \ i \,e \,B .
\end{equation}
To obtain the eigenvalues and eigenstates of the Landau Hamiltonian
(\ref{Eq:Landau_Hamiltonian}), we introduce the ladder (annihilation and creation)
operators by
\begin{equation}
 a \ = \ \sqrt{\frac{1}{2 \,e \,B}} \,\left( \Pi_x \ + \ i \,\Pi_y \right), \hspace{6mm}
 a^\dagger \ = \ \sqrt{\frac{1}{2 \,e \,B}} \,\left( \Pi_x \ - \ i \,\Pi_y \right) .
\end{equation}
They satisfy the following commutation relations : 
\begin{equation}
 [a, a^\dagger] \ = \ 1, \hspace{6mm} [a, a] \ = \ [a^\dagger, a^\dagger] \ = \ 0.
\end{equation}
The Hamiltonian then reduces to
\begin{equation}
 H \ = \ \frac{1}{2 \,m} \,\left( \Pi_x^2 \ + \ \Pi_y^2 \right) \ = \ 
 \omega \,\left( a^\dagger \,a \ + \ \frac{1}{2} \right) . \label{Eq:HO_Hamiltonian}
\end{equation}
Since the last expression is nothing but the Hamiltonian of 1-dimentional harmonic
oscillator, its eigenstates and eigenvalues are readily obtained as
\begin{equation}
 H \,|\,n \rangle \ = \ E_n \,|\,n \rangle, \hspace{6mm} \mbox{with} \hspace{6mm}
 E_n \ = \ \left( n \ + \ \frac{1}{2} \right) \,\omega ,
\end{equation} 
where
\begin{equation}
 |\,n \rangle \ = \ \frac{(a^\dagger)^n}{\sqrt{n \,!}} \,|\,0 \rangle .
\end{equation}
Actually, it is a widely-known fact that each Landau level with given $n$ is 
infinitely-degenerated. To understand this degeneracy, let us introduce the two
operators $X$ and $Y$, which have the meaning of the center of cyclotron motion : 
\begin{eqnarray}
 X \ &\equiv& \ x \ + \ \frac{v_y}{\omega} \ = \ x \ + \ \frac{1}{e \,B} \,\Pi_y, \\
 Y \ &\equiv& \ y \ - \ \frac{v_x}{\omega} \ = \ y \ - \ \frac{1}{e \,B} \,\Pi_x .
\end{eqnarray}
They satisfy the following commutation relation : 
\begin{equation}
 [X, Y] \ = \ - \,i \,\frac{1}{e \,B} .
\end{equation}
Here, we introduce another ladder operator $b$ and $b^\dagger$ by
\begin{equation}
 b \ = \ \sqrt{\frac{e \,B}{2}} \,(X \ - \ i \,Y), \hspace{6mm}
 b^\dagger \ = \ \sqrt{\frac{e \,B}{2}} \,(X \ + \ i \,Y) .
\end{equation}
It is an easy exercise to check that they satisfy the following commutation
relations : 
\begin{equation}
 [b, b^\dagger] \ = \ 1, \hspace{6mm} [b, b] \ = \ [b^\dagger, b^\dagger] \ = \ 0.
\end{equation}
Furthermore, $b$ and $b^\dagger$ commute with either of $a$ and $a^\dagger$ as
\begin{equation}
 [b, a] \ = \ [b, a^\dagger] \ = \ [b^\dagger, a] \ = \ [b^\dagger, a^\dagger] \ = \ 0.
\end{equation}

Of our particulat interest is the relation between the two orbital angular momenta,
i.e. the canonical OAM and the mechanical OAM. Since the motion of the charged particle
is confined in the $x$-$y$ plane, we have only to consider the $z$-component of the
orbital angular momenta. The canonical OAM is given by
\begin{equation}
 L_{can} \ \equiv \ x \,p_y \ - \ y \,p_x ,
\end{equation}
whereas the mechanical OAM is given by
\begin{equation}
 L_{mech} \ \equiv \ m \,(x \,v_y \ - \ y \,v_x) \ = \ x \,\Pi_y \ - \ y \,\Pi_x .
\end{equation}
In the symmetric gauge, the relation between these two OAMs is given as
\begin{equation}
 L_{can} \ = \ L_{mech} \ + \ \frac{e \,B}{2} \,(x^2 \ + \ y^2) .
\end{equation}
It is interesting to point out that the difference between the canonical and mechanical OAMs is
just given by the ``potential angular momentum" introduced in \cite{Waka10,Waka11A} : 
\begin{equation}
 L_{pot} \ = \ e \,(\bm{r} \times \bm{A})_z \ = \ \frac{e \,B}{2} \,(x^2 \ + \ y^2) ,
\end{equation}
which means that
\begin{equation}
 L_{can} \ = \ L_{mech} \ + \ L_{pot} . \label{Eq:total_OAM}
\end{equation}
As explained in \cite{Review_Waka14,Waka10,Waka11A}, 
$L_{pot}$ represents the angular momentum carried by the electromagnetic
potential, which is the external magnetic field in the present problem. 
Eq.(\ref{Eq:total_OAM}) thus means that the
canonical OAM represents the total OAM, that is, the {\it sum} of the particle OAM and the
OAM carried by the electromagnetic field. (A support to this interpretation is also
found in a recent paper \cite{GSFB14}.)

To proceed, we express $L_{mech}$ and $L_{can}$ in terms of
the ladder operators $a, a^\dagger, b$ and $b^\dagger$.  The answer is given by
\begin{eqnarray}
 L_{mech} \ &=& \ i \,(b \,a^\dagger \ - \ b^\dagger \,a) \ - \ (a^\dagger \,a \ + \ a \,a^\dagger), \\
 L_{can} \ &=& \ \frac{1}{2} \,(b^\dagger \,b \ + \ b \,b^\dagger) \ - \ \frac{1}{2} \,
 (a^\dagger \,a \ + \ a \,a^\dagger) .
\end{eqnarray}
(Remember that the Hamiltonian is already expressed as (\ref{Eq:HO_Hamiltonian}) 
only with $a$ and $a^\dagger$.)
Using the commutation relations of $a, a^\dagger, b$ and $b^\dagger$, one can easily verify
that the canonical OAM operator $L_{can}$ commutes with the Hamiltonian : 
\begin{equation}
 [L_{can}, H] \ = \ 0 ,
\end{equation}
although the mechanical OAM operator does not. This means that we can construct simultaneous
eigenstates of $H$ and $L_{can}$, which are  characterized by two harmonic oscillator quanta
$n$ and $m$ as
\begin{eqnarray}
 H \,|\,n, n + m \rangle \ &=& \ \left( n \ + \ \frac{1}{2} \right) \,\omega \,|\,n, n + m \rangle, \\
 L_{can} \,|\,n, n + m \rangle \ &=& \ m \,|\,n, n + m \rangle , 
\end{eqnarray}
where
\begin{equation}
 |\,n, m \rangle \ = \ \frac{(a^\dagger)^n \,(b^\dagger)^m}{\sqrt{n \,! \,m \,!}} \,|\,0, 0 \rangle .
\end{equation}
Here, $n$ are non-negative integers ($n = 0, 1, \cdots$) characterizing the Landau level, 
while $m$ are integers satisfying
the inequality $m \geq - \,n$.
Thus, for a fixed Landau label $n$ with the eigen-energy 
$E_n = \left( n + \frac{1}{2} \right) \,\omega$, there are infinitely many states with exactly the
same eigen-energy but different $z$-component of the canonical OAM.

To understand the physical content of the two OAMs, let us investigate the expectation value
of the canonical and mechanical OAMs in the eigenstate $|\,n, n + m \rangle$,
defined by
\begin{equation}
 \langle O \rangle \ \equiv \ \langle n, n + m \,| \,O \,|\,n, n + m \rangle. 
\end{equation}
As can be easily checked, the expectation value of the mechanical OAM becomes
\begin{equation}
 \langle L_{mech} \rangle \ = \ - \,(2 \,n \ + \ 1) ,
\end{equation}
which is independent of $m$.
The expectation value of the potential angular momentum can also be readily calculated as
\begin{equation}
 \langle L_{pot} \rangle \ = \ \frac{e \,B}{2} \,\langle x^2 \ + \ y^2 \rangle \ = \ 
 m \ + \, (2 \,n \ + \ 1) .
\end{equation}
Adding up these two quantities, we find that
\begin{equation}
 \langle L_{can} \rangle \ = \ \langle L_{mech} \rangle \ + \ \langle L_{pot} \rangle \ = \ m ,
\end{equation}
which naturally reproduces the eigenvalue of $L_{can}$ in the state
$|\,n , n + m \rangle$..

Somewhat surprisingly, the canonical OAM characterized
by the quantum number $m$ has little to do with the physical cyclotron motion
of a charge particle in the magnetic field. This is reflected in the fact that the
quantum number $m$ does not appear in the (observable) energy 
$E_n = \left( n + \frac{1}{2} \right)$ of the Landau problem, so that it is not
a direct observable. On the other hand, the expectation value of the mechanical OAM
is characterized by the Landau quantum number $n$, so that it is clearly an observable.
Remember that the eigen-energy of the Landau level $n$ is just consistent with
the Bohr-Sommerfeld quantization condition
corresponding to the semi-classical cyclotron motion.  

Undoubtedly, this noticeable difference between the two OAMs is not unrelated to
the fact that the canonical OAM is not a gauge-invariant quantity.
To confirm it, let us investigate both OAMs in a different gauge from the symmetric
gauge, for example, in the Landau gauge. The gauge transformation from the symmetric gauge
$\bm{A}_S (\bm{r}) = \frac{B}{2} \,(- \,y, x, 0)$ to the Landau gauge
$\bm{A}_L (\bm{r}) = B \,(- \,y, 0, 0)$ is given by
\begin{equation}
 \bm{A}_L (\bm{r}) \ = \ \bm{A}_S (\bm{r}) \ + \ \nabla \,\chi (\bm{r}) ,
\end{equation}
with the choice of the gauge function
\begin{equation}
 \chi (\bm{r}) \ = \ - \,\frac{B}{2} \, x \,y .
\end{equation}
Note that, in the Landau gauge, the mechanical momenta take the form
\begin{eqnarray*}
 \Pi_x \ &\equiv& \ p_x \ - \ e \,A_x \ = \ p_x \ + \ e \,B \,y , \\
 \Pi_y \ &\equiv& \ p_y \ - \ e \,A_y \ = \ p_y .
\end{eqnarray*}
The Hamiltonian is given by
\begin{equation}
 H \ = \ \frac{1}{2 \,m} \,(\Pi_x^2 \ + \ \Pi_y^2) \ = \ \frac{1}{2 \,m} \,
 \left\{ (p_x \ + \ e \,B \,y)^2 \ + \ p_y^2 \right\} .
\end{equation}
Since this Hamiltonian does not contain the coordinate $x$, its eigenfunction is given as
\begin{equation}
 \psi (x, y) \ \propto \ e^{\,i \,k_x \,x} \,\,\phi (y) ,
\end{equation}
or in more abstract form as
\begin{equation}
 |\, \psi \rangle \ = \ |\,k_x, \phi \rangle \ \equiv \ |\,k_x \rangle \,|\,\phi \rangle ,
\end{equation}
with $|\,k_x \rangle$ being the eigenstate of $p_x$ : 
\begin{equation}
 p_x \,|\,k_x \rangle \ = \ k_x \,|\,k_x \rangle .
\end{equation}
This leads to an effective Hamiltonian in the $y$-space as
\begin{equation}
 H^\prime \ = \ \frac{1}{2 \,m} \,\left\{ (k_x \ + \ e \,B \,y)^2 \ + \ p_y^2 \right\} .
\end{equation}
This is essentially the Hamiltonian of 1-dimensional Harmonic oscillator, so that
its eigenvalues and eigenfunctions are easily be written down as
\begin{equation}
 H^\prime \,|\,n \rangle \ = \ \left( n + \frac{1}{2} \right) \,|\,n \rangle .
\end{equation}
with
\begin{equation}
 |\,n \rangle \ = \ \frac{(a^\dagger)^n}{n \,!} \,|\,0 \rangle .
\end{equation}
Here
\begin{eqnarray}
 a &=& \sqrt{\frac{1}{2 \,e \,B}} \,\left\{ (k_x \ + \ e \,B \,y) \ + \ i \,p_y \right\} , \\
 a^\dagger &=& \sqrt{\frac{1}{2 \,e \,B}} \,\left\{ (k_x \ + \ e \,B \,y) \ - \ i \,p_y \right\} .
\end{eqnarray}
are the ladder operators in the Landau gauge.

To sum up, the eigenenergies and eigenstates of the original Hamiltonian are
expressed as
\begin{equation}
 |\,k_x, \phi \rangle \ = \ |\,k_x, n \rangle \ = \ |\,k_x \rangle \,|\, n \rangle .
\end{equation}
with
\begin{equation}
 \langle x \,|\, k_x \rangle \ = \ \frac{1}{\sqrt{L_x}} \,e^{\,i \,k_x \,x} .
\end{equation}
Here use the box normalization for the plane-wave in the $x$-plane
with large but finite length $L_x$.

To proceed, it is convenient to write the ladder operators in the form : 
\begin{eqnarray}
 a \ &=& \ \sqrt{\frac{1}{2 \,e \,B}} \,\left\{\, e \,B \,(y \ - \ Y) \ + \ i \,p_y \right\} , \\
 a^\dagger \ &=& \ \sqrt{\frac{1}{2 \,e \,B}} \,\left\{\, e \,B \,(y \ - \ Y) \ - \ i \,p_y \right\} .
\end{eqnarray}
with
\begin{equation}
 Y \ \equiv \ - \,\frac{k_x}{e \,B} .
\end{equation}
Here, $Y$ has the meaning of the center of cyclotron motion projected on
the $y$-axis. Using the equations
\begin{equation}
 p_y \ = \ \frac{1}{i} \,\sqrt{\frac{e \,B}{2}} \,(a \ - \ a^\dagger), \hspace{6mm}
 y \ - \ Y = \frac{1}{\sqrt{2 \,e \,B}} \,(a \ + \ a^\dagger) ,
\end{equation}
one can easily verify the following relations : 
\begin{eqnarray}
 \langle n \,|\,p_y \,|\, n \rangle \ &=& \ 0, \\
 \langle n \,|\, y \ - \ Y \,|\,n \rangle \ &=& \ 0, \\
 \langle n \,|\, (y \ - \ Y)^2 \,|\,n \rangle \ &=& \ \frac{1}{e \,B} \,(2 \, n + 1) .
\end{eqnarray}

Now, we are ready to evaluate the expectation values of the two OAM operators
in the state $|\,k_x, n \rangle$.
Note first that the expectation value of the mechanical OAM can be expressed as
\begin{eqnarray}
 \langle L_{mech} \rangle \ &=& \ \langle k_x, n \,|\, x \,\Pi_y \ - \ y \,\Pi_x \,
 |\,k_x, n \rangle \nonumber \\
 \ &=& \ \langle L_{can} \rangle \ - \ \langle L_{pot} \rangle , \label{Eq:OAM_mech}
\end{eqnarray}
with
\begin{eqnarray}
 \langle L_{can} \rangle \ &\equiv& \ \langle k_x, n \,|\,x \,p_y \ - \ y \,p_x \,
 |\,k_x, n \rangle, \\
 \langle L_{pot} \rangle \ &\equiv& \ e \,B \,\langle k_x, n \,|\,y^2 \,|\,k_x, n \rangle .
\end{eqnarray}
Using the relation
\begin{equation}
 \langle n \,|\,y^2 \,|\, n \rangle \ = \ Y^2 \ + \ \frac{1}{e \,B} \,(2 \,n + 1),
\end{equation}
we find that
\begin{equation}
 \langle L_{pot} \rangle \ = \ \frac{k_x^2}{e \,B} \ + \ (2 \,n + 1) .
\end{equation}
On the other hand, we get
\begin{eqnarray}
 \langle L_{can} \rangle \ &=& \ \langle k_x \,|\,x \,|\,k_x \rangle \,
 \langle n \,|\,p_y \,|\,n \rangle \ - \ k_x \,\langle n \,|\,y \,|\,n \rangle
 \ = \ 0 \ - \,k_x \,Y \ = \ \frac{k_x^2}{e \,B} . 
\end{eqnarray}
Here, we have used the relation
\begin{equation}
 \langle n \,|\,p_y \,|\, n \rangle \ = \ 0 .
\end{equation}
Note that, although $\langle k_x \,|\,x \,|\,k_x \rangle$ diverges in the limit
$L_x \rightarrow \infty$, this limit can be taken after using the relation
$\langle n \,|\,p_y \,|\,n \rangle = 0$, or we can keep $L_x$ large but finite value.

One sees that the expectation value of the canonical OAM operator in the Landau
gauge does not coincide with that in the symmetric gauge. (The same is
true also for the potential angular momentum operator.)
On the other hand, from Eq.(\ref{Eq:OAM_mech}), the expectation value of the
mechanical OAM operator in the Landau gauge is given by
\begin{equation}
 \langle L_{mech} \rangle \ = \ \frac{k_x^2}{e \,B} \ - \ 
 \left\{ \frac{k_x^2}{e \,B} \ + \ (2 \,n + 1) \right\} \ = \ 
 - \,(2 \,n + 1) ,
\end{equation}
which precisely reproduce the expectation value of the mechanical OAM operator
in the symmetric gauge. The expectation value of the mechanical OAM operator
is therefore gauge-independent as expected. Undoubtedly, the demonstration above
implies unphysical nature of the canonical OAM, in spite that the canonical momentum
as well as the canonical OAM are useful objects in solving the quantum
mechanical problem. (More generally speaking, the canonical momentum is a
fundamental element in the canonical formaliam of quantum theory.)
On the other hand, the mechanical OAM
is gauge-invariant and it describes the physical cyclotron motion of a charged
particle in the magnetic field. This analysis within a solvable system
clearly shows the superiority of the mechanical OAM over the canonical OAM
as a physical OAM of a charge particle under the presence of strong
magnetic field. In our opinion, it also throws a slight doubts on the
physical relevance or the observability of the canonical OAM of quarks, which
appears in the Jaffe-Manohar decomposition of the nucleon. 
In the following sections, we shall investigate this QCD problem by keeping in mind
the lesson learned from the Landau problem.

\section{The nonabelian Stokes theorem and the two types of quark OAMs in the nucleon}
\label{sec:3}

An important lesson learned from the Landau problem is that, under the
presence of strong magnetic field, one must pay the finest care to the
physical difference between the
two types of OAMs, i.e. the canonical one and the mechanical one. 
As first recognized by Burkardt \cite{Burkardt13}, the existence of the
two types of quark OAMs in the nucleon is deeply connected with the
existence of strong color-electromagnetic field inside the nucleon, which is
generated by the QCD dynamics of bound quarks and gluons.
As we shall see below, the essence of Burkartdt's observation can more transparently be
understood on the basis of the nonabelian Stokes theorem as pointed out in
a recent paper by Tiwari \cite{Tiwari15}.

The nonabelian Stokes theorem is an identity for the Wilson-loop operator
\begin{equation}
 W (C) \ = \ \mbox{Tr} \,P \,\exp \left( i \,g \,\oint_C \,d z_\mu \,A^\mu (z) \right) ,
\end{equation}
where $C$ is a closed path in the 4-dimensional space-time, $\mbox{Tr}$ stands for
the trace in color space, while $P$ does the color-space path ordering operator. 
The theorem states that \cite{FGK1981,Szczesny1987}
\begin{eqnarray}
 \mbox{Tr} \,P \,\exp \left( i \,g \,\oint_C \,d z_\mu \,A^\mu (z) \right) \ = \ 
 \mbox{Tr} \,P \exp \left( i \,g \,\int_S \,d \sigma_{\mu \nu} \,\tilde{F}^{\mu \nu} (y) \right) ,
\end{eqnarray}
where
\begin{equation}
 \tilde{F}^{\mu \nu} (y) \ = \ {\cal L} [a,y] \,F^{\mu \nu} (y) \,{\cal L} [y,a] ,
\end{equation}
with $F^{\mu \nu} = \partial^\mu A^\nu - \partial^\nu \,A^\mu - i \,g \,[A^\mu, A^\nu]$
being the field strength tensor for the nonabelian gauge field, whereas
\begin{equation}
 {\cal L} [y, x] \ = \ P \,\exp \left(i \,g \,\int_x^y \,d z_\mu \,A^\mu (z) \right)
\end{equation}
is gauge-link operator connecting the two space-times points $x$ and $y$.

We apply this theorem to the average transverse momenta of quarks in the transversely
polarized nucleon and also to the average longitudinal OAM of quarks in the
longitudinally polarized nucleon, which were investigated by Burkardt in \cite{Burkardt13}.
They are respectively defined by
\begin{eqnarray}
 \langle k_\perp^l \rangle^{\cal L} \ &=& \ \int \,d x \,\int \,d^2 \bm{b}_\perp \,
 \int \,d^2 \bm{k}_\perp \,k_\perp^l \,\rho^{\cal L} (x, \bm{b}_\perp, \bm{k}_\perp \,;\, S_\perp), 
 \label{Eq:average_momentum} \\
 \langle L^3 \rangle^{\cal L} \ &=& \ \int \,d x \,\int \,d^2 \bm{b}_\perp \,
 \int \,d^2 \bm{k}_\perp \,(\bm{b} \times \bm{k}_\perp )^3 \,\rho^{\cal L} 
 (x, \bm{b}_\perp, \bm{k}_\perp \,;\, S_\parallel) , \label{Eq:average_OAM} 
\end{eqnarray}
where $l = 1$, or $2$. The Wigner distributions $\rho^{\cal L}$ appearing in the
above equations are 5-dimensional phase space distribution defined as
\begin{eqnarray}
 \rho^{\cal L} (x, \bm{b}_\perp, \bm{k}_\perp, S) \ &=& \ 
 \frac{1}{2} \,\int \,\frac{d^2 \bm{\Delta}_\perp}{(2 \,\pi)^2} \,
 \int \,\frac{d^2 \bm{\xi}_\perp \,d \xi^-}{(2 \,\pi)^3} \nonumber \\
 \ &\times& \ e^{\,- \,i \,\bm{\Delta}_\perp \cdot \bm{b}_\perp} \,\,
 e^{\,i \,(x \,P^+ \,\xi^- \,- \,\bm{k}_\perp \cdot \bm{\xi}_\perp)} \,\,
 \langle p^\prime, s^\prime \,|\,\bar{\psi} (0) \,\gamma^+ \,
 {\cal L} [0, \xi] \,\psi (\xi) \,|\,p, s \rangle ,
\end{eqnarray}
with $P = \frac{1}{2} \,(p^\prime + p)$ and $p^\prime - p = (0, \bm{\Delta}_\perp, 0)$,
while $S = \frac{1}{2} \,(s^\prime + s)$ with $s^\prime$ and $s$ denoting the polarization
states of the final and initial nucleons. As is widely known, the Wigner distribution
generally depends on the path of the gauge link ${\cal L} [0, \xi]$ connecting
the two space-time points $\xi$ and $0$.

\begin{figure}[ht]
\begin{center}
\includegraphics[width=14cm]{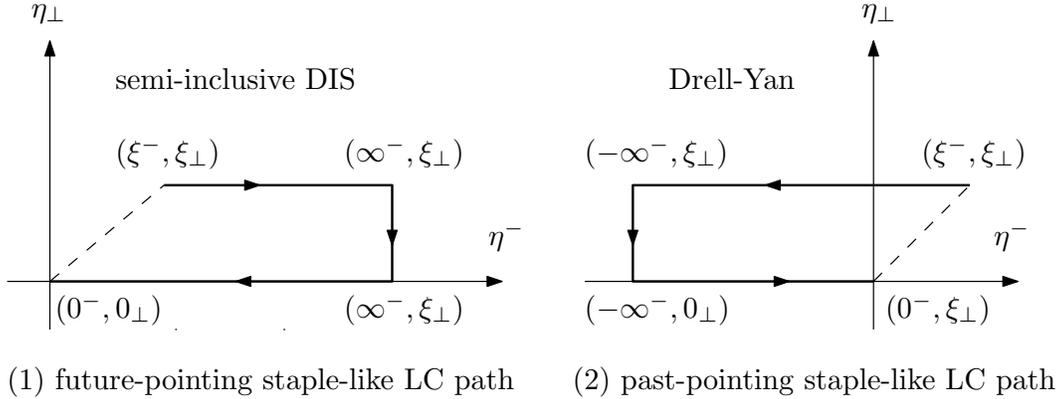}
\caption{Two gauge-link paths, which correspond to two DIS processes.}
\label{Fig:Wilson_line}
\end{center}
\end{figure}

Two physically interesting choices of the gauge-link paths are the so-called
future-pointing staple-like light-cone (LC) path denoted as ${\cal L}^{+LC}$ and the
past-pointing staple-like light-cone (LC) path denoted as ${\cal L}^{-LC}$.
 They are respectively specified as (see Fig.\ref{Fig:Wilson_line}.)
\begin{eqnarray}
 {\cal L}^{\pm LC} [0, \xi] \ &\equiv& \ 
 {\cal L}^{(st)} [0^-, \bm{0}_\perp \,;\,\pm \infty^-, \bm{0}_\perp] \nonumber \\
 \ &\,& \, \times \ 
 {\cal L}^{(st)} [\pm \infty^-, \bm{0}_\perp \,;\, \pm \infty^-, \bm{\xi}_\perp] \,
 {\cal L}^{(st)} [\pm \infty^-, \bm{\xi}_\perp \,; \, \xi^-, \bm{\xi}_\perp] ,
\end{eqnarray}
where ${\cal L}^{(st)} [\xi, \eta]$ stands for a straight-line path directly connecting
the two space-time points $\eta$ and $\xi$. (In the following, the suffix $(st)$ will
be omitted for brevity, when there is no possibility of misunderstanding.)
Remember that the above two choices of the gauge-link path
corresponds to the kinematics of semi-inclusive hadron productions 
and that of Drell-Yan processes, respectively.
In fact, a future-pointing Wilson line appears in the SIDIS processes because
the flow of color runs via an outgoing quark, whereas a past-pointing Wilson
line appears  because the flow of color runs via an incoming 
antiquark \cite{BMP06}.

\vspace{4mm}

In addition to the above two paths, also physically important is the gauge-link path
directly connecting the two space-time points $\xi$ and $0$.
Although this choice of path does not directly correspond to the kinematics of
the deep-inelastic-scattering (DIS) processes, it is nevertheless important,
since this choice in (\ref{Eq:average_momentum}) and (\ref{Eq:average_OAM}) is
known to give manifestly gauge-invariant mechanical transverse momentum and
mechanical longitudinal OAM of quarks in the nucleon \cite{JXY12}.

\vspace{4mm}
\begin{figure}[h]
\centering
\includegraphics[width=8cm]{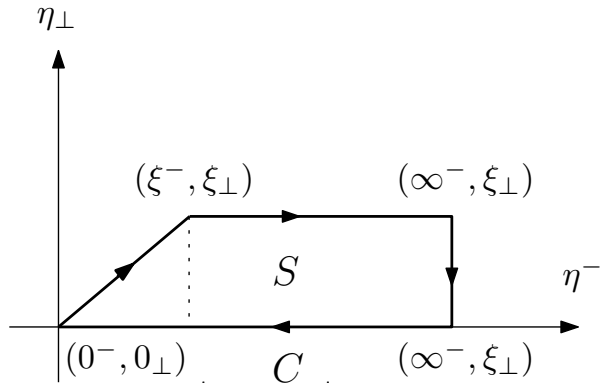}
\caption{The future-pointing staple-like LC path made closed to a loop.}
\label{Fig:Wilson_loop}
\end{figure}

Anyhow, an important fact is that, through the gauge-link path dependence of the
Wigner distribution, the average transverse momentum as well as the average
longitudinal OAM of quarks are generally path-dependent. As pointed out by
Tiwari, the reason of this path dependence can most transparently be understood
on the basis of the nonabelian Stokes theorem. Let us first consider the closed
path $C$ in the $(\eta^-, \bm{\eta}_\perp)$ plane as illustrated in Fig.2.
Because this closed gauge-link is expressed as
\begin{equation}
 {\cal L}^C [0,0] \ = \ {\cal L}^{+LC} [0,\xi] \,{\cal L}^{(st)} [\xi, 0] \ = \ 
 {\cal L}^{+LC} [0,\xi] \,\left( {\cal L}^{(st)} [0,\xi] \right)^{-1} ,
\end{equation}
we immediately obtain the following relation for the path-dependent average momenta of
quarks : 
\begin{equation}
 \langle k_\perp^l \rangle^C \ = \ \langle k_\perp^l \rangle^{+LC} \ - \ 
 \langle k_\perp^l \rangle^{straight} , \label{Eq:momentum_C1}
\end{equation}
where
\begin{eqnarray}
 \langle k_\perp^l \rangle^C \ &=& \ \int \,d x \,\int \,
 d^2 \bm{k}_\perp \,k_\perp^l \nonumber \\
 \ &\times& \ \frac{1}{2} \,\int \,\frac{d^2 \bm{\xi}_\perp \,d \xi^-}{(2 \,\pi)^3} \,
 e^{\,i \,(x \,P^+ \,\xi^- \,- \,\bm{k}_\perp \cdot \bm{\xi}_\perp)} \,
 \langle P S_\perp \,|\,\bar{\psi} (0) \,\gamma^+ \,
 {\cal L}^C [0,0] \,\psi (0) \,|\,P S_\perp \rangle , \label{Eq:momentum_C2}
\end{eqnarray}
with
\begin{equation}
 {\cal L}^C [0,0] \ = \ \mbox{Tr} \,P \,\exp \,\left( i \,g \,\oint_C \,d z_\mu \,A^\mu (z) \right) ,
\end{equation}
being the Wilson loop corresponding to the closed path $C$. 
By using the nonabelian Stokes theorem, this Wilson loop can be rewritten as
\begin{equation}
 {\cal L}^C [0,0] \ = \ \mbox{Tr} \,P \,\exp \,\left( i \,g \,\int_S \,d \sigma_{\mu \nu} (\eta) \,
 {\cal L} [0,\eta] \,F^{\mu \nu} (\eta) \,{\cal L} [\eta,0] \right) , \label{Eq:closed_path}
\end{equation}
where $S$ is an arbitrary surface with its boundary being the closed path $C$.
The physical insight obtained from the nonabelian Stokes theorem is simple
but very important. If there is no color electromagnetic flux inside the
nucleon, we would have $F^{\mu \nu} (\eta) = 0$ so that ${\cal L}^C [0,0] = 1$.
In this case, one can easily verify that the difference between the two
quantities $\langle k_\perp^l \rangle^{+LC}$ and $\langle k_\perp^l \rangle^{straight}$
vanishes identically. Conversely speaking, what generates the gauge-link path dependence
of the two definitions of the average transverse momentum of quarks is the 
existence of the color electromagnetic field inside the nucleon.

Now we can proceed as follows. First, we rewrite (\ref{Eq:momentum_C1}) with
(\ref{Eq:momentum_C2}) in the following form, 
\begin{eqnarray}
 \langle k_\perp^l \rangle^{+LC} \ - \ \langle k_\perp^l \rangle^{straight} 
 \ &=& \ 
 \int \,d x \,\int \,d^2 \bm{k}_\perp \,\,\frac{1}{2} \,
 \int \,\frac{d^2 \bm{\xi}_\perp \,d \xi^-}{(2 \,\pi)^3} \,\,
 e^{\,i \,(x \,P^+ \,\xi^- - \bm{k}_\perp \cdot \bm{\xi}_\perp)} \nonumber \\
 \ &\,& \hspace{5mm} \times \ \ \frac{1}{i} \,\frac{\partial}{\partial \xi_\perp^l} \,
 \langle P S_\perp \,|\,\bar{\psi} (0) \,\gamma^+ \,{\cal L}^C [0,0] \,
 \psi (0) \,|\,P S_\perp \rangle .
\end{eqnarray}
Here, by using the identities,
\begin{eqnarray}
 \int \,d x \,e^{\,i \,x \,P^+ \,\xi^-} \ &=& \ \frac{2 \,\pi}{P^+} \,\delta (\xi^-), \\
 \int \,d^2 \bm{k}_\perp \,e^{\,- \,i \,\bm{k}_\perp \cdot \bm{\xi}_\perp} \ &=& \ 
 (2 \,\pi)^2 \,\delta^2 (\bm{\xi}_\perp) ,
\end{eqnarray}
it reduces to
\begin{eqnarray}
 \langle k_\perp^l \rangle^{+LC} \ &-& \ \langle k_\perp^l \rangle^{straight} 
 \ = \ 
 \frac{1}{2 \,P^+} \nonumber \\
 \ &\,& \hspace{10mm} \times \ \ \left. \frac{1}{i} \,\frac{\partial}{\partial \xi_\perp^l} \,
 \langle P S_\perp \,|\,\bar{\psi} (0) \,\gamma^+ \,{\cal L}^C [0,0] \,
 \psi (0) \,|\,P S_\perp \rangle \,\right|_{\xi^- = 0, \bm{\xi}_\perp = 0} ,
\end{eqnarray}
with ${\cal L}^C [0,0]$ given by (\ref{Eq:closed_path}). 

Since the surface $S$ in the integral (\ref{Eq:closed_path}) can be taken arbitrarily as long as
its boundary is constrained to be the closed path $C$, we take it a
trapezoid in the
$(\eta^-, \eta_\perp^m)$ plane as illustrated in Fig.\ref{Fig:Wilson_loop}.
Then, we have
\begin{eqnarray}
 &\,& \int \,d \sigma_{\mu \nu} (\eta) \,{\cal L} [0,\eta] \,F^{\mu \nu} (\eta) \,
 {\cal L} [\eta, 0] \nonumber \\
 &=& \ - \ \left\{ \int_0^{\xi^-} \,d \eta^- \,\int_0^{\,(\eta^- / \xi^-) \,\xi_\perp^m} \,
 d \eta_\perp^m \,\,{\cal L} [0,\eta] \,F^{+ m} (\eta) \,{\cal L} [\eta, 0] \right. \nonumber \\
 &\,& \hspace{10mm} \ + \ \left. \int_{\xi^-}^{+ \,\infty} \,d \xi^- \,
 \int_0^{\xi_\perp^m} \,d \eta_\perp^m \,\,
 {\cal L} [0,\eta] \,F^{+ m} (\eta) \,{\cal L} [\eta, 0] \ \ \right\} .
\end{eqnarray}
This gives
\begin{eqnarray}
 &\,& \left. \frac{1}{i} \,\frac{\partial}{\partial \xi_\perp^l} \,\exp \left( 
 i \,g \,\int_S \,d \sigma_{\mu \nu} (\eta) \,{\cal L} [0,\eta] \,F^{\mu \nu} (\eta) \,{\cal L} [\eta,0] 
 \right) \,\right|_{\xi^- = 0, \bm{\xi}_\perp = 0} \nonumber \\
 &\,& \ = \ - \,g \,\delta^{l m} \,\left\{
 \int_0^{\xi^-} \,d \eta^- \,\frac{\eta^-}{\xi^-} \,
 {\cal L} \left[0^-, \bm{0}_\perp \,; \,\eta^-, \frac{\eta^-}{\xi^-} \,\xi_\perp^m \right] \,
 F^{+m} \left(\eta^-, \frac{\eta^-}{\xi^-} \,\xi_\perp^m \right) \,
 {\cal L} \left[\eta^-, \frac{\eta^-}{\xi^-} \,\xi_\perp^m \,;\, 0^-, \bm{0}_\perp \right] 
 \right. \nonumber \\
 &\,& \hspace{18mm} + \ \left. \left. \int_{\xi^-}^{+ \,\infty} \,d \eta^- \,
 {\cal L} [0^-, \bm{0}_\perp \,; \,\eta^-, \xi_\perp^m] \,F^{+m} (\eta^-, \xi_\perp^m) \,
 {\cal L} [\eta^-, \xi_\perp^m \,; \,0^-, \bm{0}_\perp] \right\} \,\right|_{\xi^- = 0, \bm{\xi}_\perp = 0} .
\end{eqnarray}
It can be shown that the first term of the above equation vanishes, while
the second term reduces to
\begin{eqnarray}
 &\,& - \,g \,\int_0^\infty \,d \eta^- \,{\cal L} [0^-, \bm{0}_\perp \,;\, \eta^-, \bm{0}_\perp] \,
 F^{+ l} (\eta^-, \bm{0}_\perp) \,{\cal L} [\eta^-, \bm{0}_\perp \,;\,0^-, \bm{0}_\perp]
 \nonumber \\
 \ &\,& \ = \ - \ g \,\int_{- \,\infty}^{+ \,\infty} \,d \eta^- \,\theta (\eta^-) \,
 {\cal L} [0,\eta^-] \,F^{+ l} (\eta^-) \,{\cal L} [\eta^-, 0] ,
\end{eqnarray}
(with $\theta (x)$ being the ordinary step function with the property $\theta (x) = 1$ for $x > 0$
and $\theta (x) = 0$ for $x < 0$) thereby being led to a simple relation
\begin{eqnarray}
 &\,& \frac{1}{i} \,\frac{\partial}{\partial \xi_\perp^l} \,\left.
 \exp \left( i \,g \,\int_S \,d \sigma_{\mu \nu} (\eta) \,{\cal L} [0,\eta] \,
 F^{\mu \nu} (\eta) \,{\cal L} [\eta, 0] \right) \right|_{\xi^- = 0, \bm{\xi}_\perp = 0} \nonumber \\
 \ &\,& \ = \ - \,g \,\int_{- \,\infty}^{+ \,\infty} \, d \eta^- \,\theta (\eta^-) \,
 {\cal L} [0,\eta^-] \,F^{+ l} (\eta^-) \,{\cal L} [\eta^-, 0] .
\end{eqnarray}
Here, we recall the fact that the average transverse momentum corresponding to the
straight-line path directly connecting $\xi$ and $0$, reduces to the mechanical
transverse momentum \cite{JXY12}
\begin{equation}
 \langle k_\perp^l \rangle^{straight} \ = \ \langle k_\perp^l \rangle_{mech} ,
\end{equation}
where
\begin{equation}
 \langle k_\perp^l \rangle_{mech} \ = \ \frac{1}{2 \,P^+} \,
 \langle P S_\perp \,|\,\bar{\psi} (0) \,\gamma^+ \,D_\perp^l (0) \,\psi (0) \,|\, P S_\perp \rangle ,
\end{equation}
with $D_\perp^l = \partial^l - \,i g \,A_\perp^l$ being the usual covariant derivative.
In this way, we eventually arrive at a key relation
\begin{eqnarray}
 &\,& \langle k_\perp^l \rangle^{+LC} \ - \ \langle k_\perp^l \rangle_{mech} \nonumber \\
 &=& \ \frac{1}{2 \,P^+} \,\langle P S_\perp \,|\,\bar{\psi} (0) \,\gamma^+ \,
 \int_{- \,\infty}^{+ \,\infty} \,d \eta^- \,( - \,\theta (\eta^-) ) \,
 {\cal L} [0,\eta^-] \,g \,F^{+l} (\eta^-) \,{\cal L} [\eta^-,0] \,\psi (0) \,
 |\,P S_\perp \rangle . \ \ \ 
\end{eqnarray}
According to Burkardt \cite{Burkardt13}, the r.h.s. of the above equation has a meaning of
final-state interaction (FSI) in the SIDIS processes.
In more detail, it represents the change of transverse momentum of the ejected
quark due to the color Lorentz force caused by the residual target.
In fact, in the LC gauge, the gauge-link along the light-cone direction becomes
unity and the relevant component of the field-strength tensor reduces to
\begin{equation}
 - \,\sqrt{2} \,g \,F^{+2} \ = \ - g \,F^{02} \ - \ g \,F^{32} \ = \ 
 g \,[\,\bm{E} \ + \ \bm{v} \times \bm{B} ]^2 ,
\end{equation}
which is nothing but the $y$-component of the color Lorentz force acting on
a particle that moves with the light velocity $\bm{v} = (0, 0, - \,1)$ in
the $- \,z$ direction \cite{Burkardt13}.

Repeating the same manipulation for the closed gauge-link
\begin{equation}
 {\cal L}^{C^\prime} [0,0] \ = \ {\cal L}^{-LC} [0,\xi] \,{\cal L}^{(st)} [\xi, 0] \ = \ 
 {\cal L}^{-LC} [0,\xi] \,\left( {\cal L}^{(st)} [0,\xi] \right)^{-1} ,
\end{equation}
containing the past-pointing staple-like LC path ${\cal L}^{-LC}$, we get
an analogous relation for $\langle k^l_\perp \rangle^{-LC}$.
Putting the two cases together, the answers can be summarized as
\begin{eqnarray}
 \langle k_\perp^l \rangle^{\pm LC} \ = \ \langle k_\perp^l \rangle_{mech}
 \ + \ \langle k_\perp^l \rangle^{\pm LC}_{int} , \label{Eq:momentum_basic}
\end{eqnarray}
with
\begin{eqnarray}
 \langle k_\perp^l \rangle^{\pm LC}_{int} &=& \frac{1}{2 \,P^+} \,\,
 \int_{- \,\infty}^{+ \,\infty} \,\,\left( \mp \,\theta (\pm \eta^- ) \right)
 \nonumber \\
 &\,& \hspace{5mm} \times \ \langle P S_\perp \,|\,\bar{\psi} (0) \,\gamma^+ \,
 {\cal L} [0,\eta^-] \,g \,F^{+l} (\eta^-) \,{\cal L} [\eta^-,0] \,\psi (0) \,
 |\,P S_\perp \rangle . \ \ \ \ \ \  
\end{eqnarray}
Here, $\langle k^l_\perp \rangle^{+ LC}_{int}$ represents the FSI in the SIDIS
processes, while $\langle k^l_\perp \rangle^{- LC}_{int}$ does the
ISI (initial-state interaction) in the Drell-Yan
processes. Note that Eq.(93) with (94) precisely reproduces the relations
derived by Burkardt with a different method. 
We point out that essentially the same relations were also obtained
by Boer et al. although in a somewhat different form \cite{BMP03}. 
(See also \cite{JY02,BJY04}.) 
To verify it, we recall a mathematical identities
\begin{equation}
 \int_{- \,\infty}^{+ \,\infty} \,d x \,\,\frac{i}{x \mp i \,\epsilon} \,e^{\,i \,\lambda x} \ = \ 
 \mp \,2 \,\pi \,\theta (\pm \,\lambda) . \label{Eq:math_identity}
\end{equation}
Using them, the above FSI or ISI term can also be expressed in the form
\begin{eqnarray}
 &\,& \langle k_\perp^l \rangle^{\pm LC}_{int}
 \ = \ \frac{1}{2 \,P^+} \,\int \,\frac{d x}{2 \,\pi} \,
 \int_{- \,\infty}^{+\,\infty} \,d \eta^- \,\frac{i}{x \mp i \,\epsilon} \,\,
 e^{\,i \,x \,P^+ \,\eta^-} \nonumber \\
 &\,& \hspace{28mm} \times \ \langle P S_\perp \,|\,\bar{\psi} (0) \,\gamma^+ \,
 {\cal L} [0,\eta^-] \,g \,F^{+l} (\eta^-) \,{\cal L} [\eta^-,0] \,\psi (0) \,
 |\,P S_\perp \rangle .
\end{eqnarray}
which corresponds to the 2nd term of r.h.s. of Eq.(4) in \cite{BMP03}. 

A similar analysis can also be carried out for the average longitudinal OAM of quarks
in the longitudinally polarized nucleon. The answer is given as
\begin{eqnarray}
 \langle L^3 \rangle^{\pm LC} \ = \ \langle L^3 \rangle_{mech}
 \ + \ \langle L^3 \rangle^{\pm LC}_{int} , \label{Eq:OAM_basic}
\end{eqnarray}
where
\begin{eqnarray}
 \langle L^3 \rangle_{mech} \ &=& \ {\cal N} \,\int \,d^2 \bm{b}_\perp \,
 \epsilon_\perp^{ij} \,b_\perp^i 
 \ \langle P S_\parallel \,|\,\bar{\psi} (0^-,\bm{b}_\perp) \,\gamma^+ \,
 \frac{1}{i} \,D_\perp^j (0^-,\bm{b}_\perp) \,\psi (0^-,\bm{b}_\perp) \,| \,
 P S_\parallel \rangle , \ \ \ \  
\end{eqnarray}
with ${\cal N} = 1 \,/\,(2 \,P^+ \,\int \,d^2 \bm{b}_\perp)$
is the manifestly gauge-invariant mechanical OAM, while
\begin{eqnarray}
 &\,& \langle L^3 \rangle^{\pm LC}_{int}
 \ = \ {\cal N} \,\int \,d^2 \bm{b}_\perp \,\,\epsilon_\perp^{i j} \,b_\perp^i
 \int_{- \,\infty}^{+ \,\infty} \,d \eta^- \,
 ( \mp \,\theta( \pm \,\eta^-)) \,\, \nonumber \\
 &\times& \, \langle P S_\parallel \,|\,\bar{\psi} (0^-, \bm{b}_\perp) \,\gamma^+ \,
 {\cal L} [0^-, \bm{b}_\perp \,;\, \eta^-, \bm{b}_\perp] \,g \,F^{+j} (\eta^-, \bm{b}_\perp) \,
 {\cal L} [\eta^-, \bm{b}_\perp, 0^-  \,;\, \bm{b}_\perp] \,\psi (0^-, \bm{b}_\perp) \,
 |\, P S_\parallel \rangle , \ \ \ \ \ \ \ \ 
\end{eqnarray}
is the FSI or ISI term.
(Here, $\epsilon^{ij}_\perp \,(i,j = 1,2)$ is the antisymmetric tensor in the transverse
plane with the convention $\epsilon^{12}_\perp = + \,1$.) 
Again, this precisely reproduces the relation derived by Burkardt \cite{Burkardt13}.
Alternatively, by using the identities (\ref{Eq:math_identity}), the FSI or ISI term can also be
expressed in the following form : 
\begin{eqnarray}
 &\,& \langle L^3 \rangle^{\pm LC}_{int}
 \ = \ {\cal N} \,\int \,\frac{d x}{2 \,\pi} \,
 \int \,d^2 \bm{b}_\perp \,\,\epsilon_\perp^{i j} \,b_\perp^i
 \int_{- \,\infty}^{+ \,\infty} \,d \eta^- \,
 \frac{i}{x \pm i \,\epsilon} \,\,e^{\,i \,x \,P^+ \,\eta^-} 
 \nonumber \\
 &\times& \, \langle P S_\parallel \,|\,\bar{\psi} (0^-, \bm{b}_\perp) \,\gamma^+ \,
 {\cal L} [0^-, \bm{b}_\perp \,;\, \eta^- , \bm{b}_\perp] \,g \,F^{+j} (\eta^-, \bm{b}_\perp) \,
 {\cal L} [\eta^-, \bm{b}_\perp \,;\, 0^- ,\bm{b}_\perp] \,\psi (0^-, \bm{b}_\perp) \,
 |\, P S_\parallel \rangle . \ \ \ \ \ \ \ \ 
\end{eqnarray}
The physical interpretation of the above relations are essentially the same
as the average transverse
momentum case. The term $\langle L^3 \rangle^{+LC}_{int}$ represents
the FSI in the SIDIS processes, while $\langle L^3 \rangle^{-LC}_{int}$ 
does the ISI in the Drell-Yan
processes. The only change from the previous case is that the role of
color Lorentz force is now replaced by the torque of it given by
\begin{equation}
 T^z \ = \ g \,\left[\, \bm{b}_\perp \times \,
 (\bm{E} \ + \ \bm{v} \times \bm{B}) \right]^3 .
\end{equation}

\section{On the idea of physical component of the gauge field}
\label{sec4}

It is important to recognize the fact that the theoretical formulation
so far is absolutely independent of the issue of a proper definition of the
the physical component of the gauge field, 
which brought about a lot of controversies in the nucleon spin decomposition
problem.
Note that each term on the r.h.s. of the relations (\ref{Eq:momentum_basic})
and (\ref{Eq:OAM_basic}) have clear and unambiguous physical meaning.
Namely, the 1st terms of (\ref{Eq:momentum_basic}) and (\ref{Eq:OAM_basic})
represent the manifestly gauge-invariant
mechanical momentum and the mechanical OAM, respectively,
whereas the 2nd terms in the same equations stand for the FSI in the SIDIS
processes or the ISI in the Drell-Yan processes. 
Unfortunately, there is some delicacy in the interpretation of the l.h.s.
In particular, if one wants to relate Eq.(\ref{Eq:OAM_basic}) to the problem of
gauge-invariant complete decomposition of the nucleon spin,
one cannot stay out of
the idea of physical component of the gauge field $A^\mu_{phys}$. 
According to Hatta \cite{Hatta12}, the original proposal for $A^\mu_{phys}$ 
by Chen et al. \cite{Chen08,Chen09}
based on the nonabelian generalization of the transverse component of
the photon field is not acceptable, because it does not correspond to
observable decomposition of the nucleon spin brobed by DIS measurements. 
Instead, he proposed three
candidates for the proper definition of $A^\mu_{phys}$, given as
\begin{eqnarray}
 A^j_{phys} (0) \ &=& \ \int_{- \,\infty}^{+ \,\infty} \,d \eta^- \,\,
 \left( - \,\theta (+ \,\eta^-) \right) \,{\cal L} [0, \eta^-] \,
 F^{+ j} (\eta^-) \,{\cal L} [\eta^-, 0] \nonumber \\
 \ &=& \ \int \,\frac{d x}{2 \,\pi} \,\int_{- \,\infty}^{+ \,\infty} \,
 d \eta^- \,\,\frac{i}{x - i \,\epsilon} \,\,e^{i \,x \,P^+ \, \eta^-} \,\, 
 {\cal L} [0, \eta^-] \,
 F^{+ j} (\eta^-) \,{\cal L} [\eta^-, 0] , \label{Eq:post_form}
\end{eqnarray}
which will be called the post-form here, or as
\begin{eqnarray}
 A^j_{phys} (0) \ &=& \ \int_{- \,\infty}^{+ \,\infty} \,d \eta^- \,\,
 \left( + \,\theta (- \,\eta^-) \right) \,{\cal L} [0, \eta^-] \,
 F^{+ j} (\eta^-) \,{\cal L} [\eta^-, 0] \nonumber \\
 \ &=& \ \int \,\frac{d x}{2 \,\pi} \,\int_{- \,\infty}^{+ \,\infty} \,
 d \eta^- \,\,\frac{i}{x + i \,\epsilon} \,\,e^{i \,x \,P^+ \, \eta^-} \,\, 
 {\cal L} [0, \eta^-] \,
 F^{+ j} (\eta^-) \,{\cal L} [\eta^-, 0] , \label{Eq:prior_form}
\end{eqnarray}
called the prior-form, or
\begin{eqnarray}
 A^j_{phys} (0) \ &=& \ - \,\frac{1}{2} \int_{- \,\infty}^{+ \,\infty} \,d \eta^- \,\,
 \epsilon(\eta^-) \,{\cal L} [0, \eta^-] \,
 F^{+ j} (\eta^-) \,{\cal L} [\eta^-, 0] \nonumber \\
 \ &=& \ \int \,\frac{d x}{2 \,\pi} \,\int_{- \,\infty}^{+ \,\infty} \,
 d \eta^- \,\,P \,\frac{i}{x} \,\,e^{i \,x \,P^+ \, \eta^-} \,\, 
 {\cal L} [0, \eta^-] \,
 F^{+ j} (\eta^-) \,{\cal L} [\eta^-, 0] , \label{Eq:PV_form}
\end{eqnarray}
called the principle-value form.
As pointed out by Hatta, for any of the above three choices, the parity and
time-reversal (PT) symmetries ensures that the FSI and ISI terms
in (\ref{Eq:OAM_basic}) precisely coincide and reduce
to the following form \cite{Hatta12}
\begin{equation}
 \langle L^3 \rangle^{+ LC}_{int} \ = \ 
 \langle L^3 \rangle^{- LC}_{int} \ = \ {\cal N} \,\int \,d^2 \bm{b}_\perp \,\,
 \epsilon^{ij}_\perp \,\,b^i_\perp \,\,\langle P S_\parallel \,|\,
 \bar{\psi} (\bm{b}_\perp) \,\gamma^+ \,g \,A^j_{phys} (\bm{b}_\perp) \,
 \psi (\bm{b}_\perp) \,|\, P S_\parallel \rangle ,
\end{equation}
which can be identified with the so-called potential angular momentum term
$\langle L^3 \rangle_{pot}$ according to the terminology in \cite{Waka10,Waka11A}. 
Inserting it into (98), we therefore get the relation
\begin{equation}
 \langle L^3 \rangle^{\pm LC} \ = \ \langle L^3 \rangle_{mech} \ + \ 
 \langle L^3 \rangle_{pot}  .
\end{equation}
Here, the sum of the mechanical OAM and the potential OAM reduces to
\begin{eqnarray}
 \langle L^3 \rangle_{``can"} \ &=& \ {\cal N} \,\int \, d^2 \bm{b}_\perp \,\,
 \epsilon^{ij}_\perp \,b^i_\perp \nonumber \\
 \ &\,& \ \times \ \langle P S_\parallel \,|\, \bar{\psi} (0^-, \bm{b}_\perp) \,
 \gamma^+ \,\frac{1}{i} \,D^j_{pure,\perp} (0^-, \bm{b}_\perp) \,
 \psi (0^-, \bm{b}_\perp) \,|\, P S_\parallel \rangle , \label{Eq:canonical_OAM}
\end{eqnarray}
with the definition of the so-called pure-gage covariant derivative as
\begin{equation}
 D^j_{pure, \perp} (0^-,\bm{b}_\perp) \ = \ \partial^j_\perp \ - \ g \,
 A^j_{pure,\perp} (0^-, \bm{b}_\perp) .
\end{equation}
Eq.(\ref{Eq:canonical_OAM}) is nothing but the gauge-invariant canonical OAM.
In this way, the average longitudinal OAM defined through the Wigner
distribution with the future-pointing LC path as well as with the
past-pointing LC path just coincide and both reduce to the gauge-invariant
canonical OAM 
\begin{equation}
 \langle L^3 \rangle^{+ LC} \ = \ 
 \langle L^3 \rangle^{- LC} \ = \ \langle L^3 \rangle_{``can"} ,
\end{equation}
which is physically equivalent to the canonical OAM appearing in the
Jaffe-Manohar decomposition of the nucleon spin \cite{JM1990,BJ1999}.

As emphasized in our previous paper \cite{Waka15}, however, the situation is considerably
different for the case of average transverse momentum of quarks in the
transversally-polarized nucleon. 
In fact, if we adopt the post-form definition (\ref{Eq:post_form}) of 
$A^l_{phys}$, the average
transverse momentum corresponding to the SIDIS processes reduces to
\begin{equation}
 \langle k^l_\perp \rangle^{+ LC} \ = \ 
 \langle k^l_\perp \rangle_{mech} \ + \ 
 \frac{1}{2 \,P^+} \,
 \langle P S_\perp \,|\,\bar{\psi} (0) \,\gamma^+ \,g \,A^l_{phys} (0) \,
 \psi (0) \,|\, P S_\perp \rangle ,
\end{equation}
which formally takes the form of gauge-invariant canonical momentum.
On the other hand, if we use the prior-form definition (\ref{Eq:prior_form}) 
of $A^l_{phys}$, the average transverse momentum corresponds to the 
Drell-Yan processes becomes
\begin{equation}
 \langle k^l_\perp \rangle^{- LC} \ = \ 
 \langle k^l_\perp \rangle_{mech} \ + \
 \frac{1}{2 \,P^+} \,
 \langle P S_\perp \,|\,\bar{\psi} (0) \,\gamma^+ \,g \,A^l_{phys} (0) \,
 \psi (0) \,|\, P S_\perp \rangle ,
\end{equation}
which also takes the form of gauge-invariant canonical momentum.
However, we already know the fact that the average transverse momentum
corresponding to the SIDIS processes and that corresponds to the Drell-Yan
processes have opposite signs \cite{Collins02} 
\begin{equation}
 \langle k^l_\perp \rangle^{- LC} \ = \ - \,\langle k^l_\perp \rangle^{+ LC} .
\end{equation}
This means that, at least for the average transverse momentum case,
neither of the post-form definition nor the prior-form definition of $A^l_{phys}$
is acceptable as a concept with universal or process-independent meaning. 

An important lesson learned from the above consideration is that, while
it is certainly true that the gauge-link structure of the average transverse momentum
as well as the average longitudinal OAM is determined by the kinematics of DIS
processes, the definition of the physical component $A^l_{phys}$ still has
some sort of arbitrariness. As pointed out in \cite{Waka15}, the most natural
choice of $A^l_{phys}$, which holds universally in both the average transverse
momentum case and the average longitudinal OAM case,
would be to use the principle-value prescription for $A^l_{phys}$ given
by (\ref{Eq:PV_form}).
In fact, the principle-value prescription for avoiding $1/x$ type 
singularity of the parton distributions is nothing uncommon \cite{KT1999}.
It is widely used in other situations, too. Especially relevant to our present
problem is the definition of the longitudinally polarized gluon distribution.

Let us start here with the popular definition of the longitudinally polarized
gluon distribution given in the paper by Manohar \cite{Manohar1990,Manohar1991}
(see also \cite{CS1982,KT1999})
\begin{eqnarray}
 x \,\Delta g (x) \ &=& \ \frac{i}{4 \,P^+} \,\int \,\frac{d \xi^-}{2 \,\pi} \,\,
 e^{\,i \,x \,P^+ \,\xi^-} \nonumber \\
 &\,& \hspace{8mm} \times \ \left\{ \langle P S_\parallel \,|\,
 \tilde{F}^{+, a}_\lambda (0) \,{\cal L}_a^b [0, \xi^-] \,
 F^{+ \lambda}_b (\xi^-) \,|\, P S_\parallel \rangle \right. \nonumber \\
 &\,& \hspace{10mm} - \ \left. \langle P S_\parallel \,|\,
 \tilde{F}^{+, a}_\lambda (\xi^-) \,{\cal L}_a^b [\xi^-, 0] \,
 F^{+ \lambda}_b (0) \,|\, P S_\parallel \rangle \right\} , \label{Eq:lpol_gluon_dist}
\end{eqnarray}
where ${\cal L}_a^b [0, \xi^-]$ represents the gauge-link in the adjoint
representation.
Using the gauge-link in the fundamental representation, the same quantity
can also be expressed as
\begin{eqnarray}
 x \,\Delta g (x) \ &=& \ \frac{i}{4 \,P^+} \,\int \,\frac{d \xi^-}{2 \,\pi} \,\,
 e^{\,i \,x \,P^+ \,\xi^-} \nonumber \\
 &\,& \hspace{8mm} \times \ \left\{ \langle P S_\parallel \,|\,
 2 \,\mbox{Tr} \,(\tilde{F}^+_\lambda (0) \,{\cal L} [0, \xi^-] \,
 F^{+ \lambda} (\xi^-) \,{\cal L} [\xi^-,0] ) 
 \,|\, P S_\parallel \rangle \right. \nonumber \\
 &\,& \hspace{10mm} - \ \left. \langle P S_\parallel \,|\,
 2 \,\mbox{Tr} \,(\tilde{F}^+_\lambda (\xi^-) \,{\cal L} [\xi^-, 0] \,
 F^{+ \lambda} (0) \,{\cal L} [0, \xi^-] ) \,|\, P S_\parallel \rangle \right\} , 
\end{eqnarray}
Rewriting the 2nd term by utilizing the translational invariance together
with the variable change $\xi^- \rightarrow - \,\xi^-$, one can rewrite the
above equation as
\begin{eqnarray}
 x \,\Delta g(x) \ &=& \ \frac{i}{4 \,P^+} \,\int \,\frac{d \xi^-}{2 \,\pi} \,
 \left( e^{\,i \,x \,P^+ \,\xi^-} \ - \ e^{\,- \,i \,x \,P^+ \,\xi^-} \right)
 \nonumber \\
 &\times& \ \langle P S_\parallel \,|\,
 2 \,\mbox{Tr} \,(\tilde{F}^+_\lambda (0) \,{\cal L} [0, \xi^-] \,
 F^{+ \lambda} (\xi^-) \,{\cal L} [\xi^-,0] ) 
 \,|\, P S_\parallel \rangle .
\end{eqnarray}
Since the above expression shows that the distribution $\Delta g(x)$
has $1/x$ type singularity, the 1st moment of $\Delta g(x)$ have a danger
of being dependent on how to avoid this singularity. Fortunately, 
we do not need to worry about it \cite{Hatta11}. 
As is clear from the consideration
of the average transverse momentum as well as the average
longitudinal OAM, physics-motivated choices would be given by the
replacements
\begin{equation}
 \frac{1}{x} \ \rightarrow \ \frac{1}{x \mp i \,\epsilon},
\end{equation}
which correspond to the post- and prior-form prescriptions respectively
relevant for the DIS processes and the Drell-Yan processes.
However, because of the identity
\begin{eqnarray}
 &\,& \frac{1}{x \mp i \,\epsilon} \,\left( e^{\,i \,x \,P^+ \,\xi^-} \ - \ 
 e^{\,- \,x \,P^+ \,\xi^-} \right) \nonumber \\
 &=& \ P \,\frac{1}{x} \,\left( e^{\,i \,x \,P^+ \,\xi^-} \ - \ 
 e^{\,- \,x \,P^+ \,\xi^-} \right) \ \pm \ i \,\pi \,\delta (x) \,
 \left( e^{\,i \,x \,P^+ \,\xi^-} \ - \ 
 e^{\,- \,x \,P^+ \,\xi^-} \right) \nonumber \\
 &=& \ P \,\frac{1}{x} \,\left( e^{\,i \,x \,P^+ \,\xi^-} \ - \ 
 e^{\,- \,x \,P^+ \,\xi^-} \right) ,
\end{eqnarray}
only the principle-value parts survive in both cases. We thus obtain
\begin{eqnarray}
 \Delta g (x) \ &=& \ \frac{1}{4 \,P^+} \,\int \,\frac{d \xi^-}{2 \,\pi} \,
 P \,\,\frac{i}{x} \,\left( e^{\,i \,x \,P^+ \,\xi^-} \ - \ 
 e^{\,- \,x \,P^+ \,\xi^-} \right)  \nonumber \\
 &\,& \hspace{8mm} \times \ \langle P S_\parallel \,|\,
 2 \,\mbox{Tr} \,(\tilde{F}^+_\lambda (0) \,{\cal L} [0, \xi^-] \,
 F^{+ \lambda} [\xi^-, 0] ) \,|\, P S_\parallel \rangle .
\end{eqnarray}
This ensures that the longitudinally polarized gulon distribution measured
in the DIS processes and that in the Drell-Yan processes are just the
same \cite{BJY04}. 
Clearly, this is related to the PT-even nature of the longitudinally polarized
gluon distribution defined  by (\ref{Eq:lpol_gluon_dist}). 
Now, by using the identity
\begin{equation}
 \int_{- \,\infty}^{+ \,\infty} \,d x \,P \,\frac{i}{x} \,
 \left( e^{\,i \,x \,P^+ \,\xi^-} \ - \ e^{\,- \,x \,P^+ \,\xi^-} \right)
 \ = \ - \,2 \,\pi \,\epsilon (\xi^-) ,
\end{equation}
the 1st moment of $\Delta g(x)$ can be written as
\begin{equation}
 \int \,\Delta g(x) \,d x \ = \ - \,\frac{1}{4 \,P^+} \,
 \int_{- \,\infty}^{+ \,\infty} \,\epsilon (\xi^-) \,
 \langle P S_\parallel \,|\,
 2 \,\mbox{Tr} \,(\tilde{F}^+_\lambda (0) \,{\cal L} [0, \xi^-] \,
 F^{+ \lambda} [\xi^-, 0] ) \,|\, P S_\parallel \rangle .
\end{equation}
We recall that this is just the form given in the paper \cite{Jaffe1996}
by Jaffe.
Now, if we introduce the physical component of the gluon field by the equation,
\begin{eqnarray}
 A^\lambda_{phys} (0) \ &=& \ - \,\frac{1}{2} \,\int_{- \,\infty}^{+ \,\infty} \,
 d \xi^- \,\epsilon (\xi^-) \,{\cal L} [0, \xi^-] \,F^{+ \lambda} \,(\xi^-) \,
 {\cal L} [\xi^-, 0]  \nonumber \\
 \ &=& \ \int \,\frac{d x}{2 \,\pi} \,\int_{- \,\infty}^{+ \,\infty} \,
 P \,\frac{i}{x} \,\,e^{\,i \,x \,P^+ \,\xi^-} \,\,{\cal L} [0, \xi^-] \,
 F^{+ \lambda} \,{\cal L} [\xi^-,0] ,
\end{eqnarray}
the 1st moment of $\Delta g(x)$ just reduces to the familiar form
\begin{equation}
 \int \,\Delta g(x) \,d x \ = \ \frac{1}{2 \,P^+} \,
 \langle P S_\parallel \,|\, 2 \,\mbox{Tr} \,(\tilde{F}^+_\lambda (0) \,
 A^\lambda_{phys} (0) ) \,|\, P S_\parallel \rangle .
\end{equation}
%
%
%

In any case, we confirm that, once we define the
physical component of the gluon field by the equation (\ref{Eq:PV_form}), 
a gauge-invariant complete decomposition of the nucleon spin including
the gluon intrinsic spin term is possible. An delicate
question is whether it is gauge-invariant decomposition in a standard
sense.
From a formal standpoint, the r.h.s. of the definition (105)
for the physical component of the gluon looks
completely gauge-invariant, since it contains only
the field-strength tensor.
Furthermore, although this definition is motivated by
the LC gauge, it does not prevent us from working in
other gauges including the covariant gauges like the
Feynman gauge.
However, we also know that this definition of the physical component is
path-dependent and there are many indications
that the path-dependence after all means 
gauge-dependence \cite{Waka13,Belinfante1962,RS1965,Yang1985}.
Lorc\'{e} argued that the above definition of the
physical component is gauge-invariant but it is not
invariant under what-he-call the St\"{u}ckelberg
transformation \cite{Lorce14,Lorce13A,Lorce13B}.
According to him, if some quantity is gauge-invariant
but St\"{u}ckelberg-variant, such a quantity is said to
have only weak gauge-invariance. The gauge-invariant
canonical quark OAM is a typical of such quantities.
On the other hand, the mechanical quark OAM is St\"{u}ckelberg-invariant
as well as gauge-invariant.
Such a quantity is said to have strong gauge-invariance.
Admitting the existence of two forms of gauge symmetry,
an immediate question is the relation with the gauge principle of
physics, especially the relation between the observability and
the two types of gauge-symmetry.
Lorc\'{e} argued that strong
form of gauge symmetry is a sufficient condition of
observability but it is not a necessary condition.
The weak form of gauge-invariance is enough for observability.
Based on these considerations, he proposed to classify
measurable quantities into two categories as follows \cite{Lorce14} : 

\begin{itemize}
\item {\it Observables}, which are gauge-invariant quantities
in a strong sense ; 

\item {\it Quasi-observables}, which are gauge-invariant
quantities in a weak sense.
\end{itemize}

When he refers to quasi-observables, what are in his mind
are principly the parton distribution functions (PDFs).
To provide an supplementary explanation, 
we first recall that the nucleon structure
functions are genuine observables, because they appear
directly in the cross section formulas of DIS reactions.
On the other hand, the PDFs are not, since they are
theoretical concepts, which generally depend on the
factorization scheme within the framework of the perturbative
QCD.
Despite this theoretical-scheme dependence, the PDFs are
approximately (i.e. at the leading order of twist expansion)
equal to the corresponding structure functions.
In this sense, the PDFs are sometimes called quasi-observables.

One might think that the above classification of observable
is roughly to the point.
However, there remains some question for admitting it as a general rule.
In fact, according to Lorc\'{e}, the standard transverse-longitudinal
decomposition (or the Helmholtz decomposition) of the photon field
is also gauge-invariant but St\"{u}ckelberg-variant. 
The transverse-longitudinal decomposition is therefore gauge-invariant
only in a weak sense. As we have repeatedly emphasized, the reason
why the transverse-longitudinal decomposition has only weak
gauge-invariance can be explained by using more familiar concept of
physics. As far as we are working in a fixed Lorentz frame of reference,
there is no doubt that the transverse (or physical) component of the
photon is gauge-invariant \cite{VanEN94A,VanEN94B,CDG1989}.
Still, this invariance cannot be a strong
gauge-invariance, because the concept of transversality is
necessarily Lorentz-frame dependent. Importantly, however,
the measurement of the spin and OAM of the photon is carried out
in a prescribed Lorentz frame by making use of interactions with atoms.
Thus, even though the spin and OAM decomposition of the photon
is only weakly gauge-invariant, the several concrete experiments
carries out in the past definitely shows that they are genuine
observables not quasi-observables \cite{Beth1936,ABSW1992}. 
In our opinion, what ensures
the observability of a given quantity is whether there is an
external current or a probe that couples to the quantity in question.
A typical is electroweak current, which can be used to probe the
internal electroweak structure of hadrons.
In the photon spin and OAM measurement, interactions with atoms
play the role of external probes.
Turning back to the general rule of Lorc\'{e}, the canonical OAMs of quarks
as well as the gluon spin are quasi-observables, not because they are
weakly gauge-invariant quantities. This is obvious from the fact that
even the manifestly gauge-invariant OAM of quarks, which is related
to the GPDs, is also a quasi-observable. 
The quasi-observability is rather related to the fact that we are dealing
with the bound state not free photons and that, for extracting the
information on the internal quark-gluon structure of the nucleon,
we need a special theoretical framework of perturbative QCD.

To sum up, we agree that the observability does not necessarily require
strong gauge-invariance. The weak gauge-invariance is enough for
observability. Still, it would not be so easy to make a simple and
clear-cut statement on the relation between the observability and the
weak-gauge invariance. A safe statement at the present moment is
whether a given weakly gauge-invariant quantity is observable or not
can be judged only through efforts to finding out an appropriate
method of observation.
Anyhow, with the understanding gained from the above
general consideration in mind, we compare the following
4 decompositions of the nucleon spin.
They are the Ji decomposition (I) \cite{Ji1997,Ji1998}
\begin{equation}
 \frac{1}{2} \ = \ J^q \ + \ J^G ,
\end{equation}
the Ji decomposition (II) \cite{Ji1997}
\begin{equation}
 \frac{1}{2} \ = \ L^q_{mech} \ + \ \frac{1}{2} \,\Delta \Sigma \ + \ J^G ,
\end{equation}
the mechanical decomposition proposed in \cite{Waka10,Waka11A}
\begin{equation}
 \frac{1}{2} \ = \ L^q_{mech} \ + \ \frac{1}{2} \,\Delta \Sigma \ + \ 
 L^G_{mech} \ + \ \Delta G , \label{Eq:mechanical_decomposition}
\end{equation}
and the canonical decomposition, which is equivalent to the Jaffe-Manohar
decomposition \cite{JM1990,BJ1999}
\begin{equation}
 \frac{1}{2} \ = \ L^q_{``can"} \ + \ \frac{1}{2} \,\Delta \Sigma \ + \ 
 L^G_{``can"} \ + \ \Delta G , \label{Eq:canonical_decomposition}
\end{equation}
where the quark and gluon OAMs in the decomposition 
(\ref{Eq:mechanical_decomposition}) and the canonical decomposition
(\ref{Eq:canonical_decomposition}) are related by the follwowing equations.
\begin{eqnarray}
 L^q_{``can"} \ &=& \ L^q_{mech} \ + \ L_{pot}, \\
 L^G_{``can"} \ &=& \ L^G_{mech} \ - \ L_{pot}.
\end{eqnarray}
Among these 4 decomposition, manifestly gauge-invariant decompositions
are the first two. The last two decompositions, which provide us with
complete decompositions of the nucleon spin, requires the
concept of the physical component of the gluon field.

As is widely known, the total angular momenta of quarks and gluons can be
related to the 2nd moments of the generalized parton distributions (GPDs)
$H^{q/G} (x,\xi,t)$ and $E^{q/G} (x,\xi,t)$, or
equivalently the forward limits of the so-called generalized (or
gravitational) form factors
$A^{q/G} (t)$ and $B^{q/G} (t)$ as \cite{Ji1997,Ji1998}
\begin{eqnarray}
 J^q \ &=& \ \frac{1}{2} \,\int \,d x \,\, x \,\,
 \left[ H^q (x,0,0) \ + \ E^q (x,0,0) \right] \,\,\ = \ \frac{1}{2} \,\,
 \left[ A^q (0) \ + \ B^q (0) \right] , \\
 J_G \ &=& \ \frac{1}{2} \,\int \,d x \,\,x \,\,
 \left[ H^G (x,0,0) \ + \ E^G (x,0,0) \right] \ = \ \frac{1}{2} \,\,
 \left[ A^G (0) \ + \ B^G (0) \right] .
\end{eqnarray}
Although the GPDs $H^{q/G} (x,\xi,t)$ and $E^{q/G} (x,\xi,t)$ are quasi-observables
just like the PDFs, the gravitational form factors $A^{q/G} (t)$
and $B^{q/G} (t)$ can in principle be extracted frame-independently by
carrying out gedanken graviton-nucleon scatterings experiments, 
so that $J^q$ and $J^G$ may be
thought of as genuine observables as emphasized in \cite{SW00}.
(Of course, the graviton-nucleon scattering measurement is practically
impossible.)
Turning to the Ji decomposition (II), the quark spin term $\Delta \Sigma$
is usually believed to be observable. To be more strict, it is a
quasi-observable, since it is just the 1st moment of the longitudinally
polarized distribution function of quarks.
In fact, the definition of $\Delta \Sigma$ is known to be factorization-scheme
dependent. The two popular choices of factorization schemes are
the standard $\overline{\mbox{MS}}$
scheme and the so-called Adler-Bardeen (AB) scheme \cite{AEL1995}.
However, the current
understanding is that there is no compelling reason to choose the AB
scheme, which breaks gauge-invariance at the cost of chiral symmetry.
Once the $\overline{\mbox{MS}}$ scheme is chosen, $\Delta \Sigma$
can be identified with the forward limit of the flavor-singlet axial
form factor of the nucleon, which will be extracted in the near-future
measurements of the neutrino-nucleon scatterings \cite{PT14}. 
We may then be
able to say that the Ji decomposition (II) is also an observable decomposition.

As repeatedly emphasized, the last two gauge-invariant complete
decompositions of the nucleon spin requires the idea of the physical
component of the gluon field. Still, we emphasize that
there is a big difference between
these two decompositions from the observational point of view.
The gluon spin term $\Delta G$ in the mechanical decomposition
is certainly a quasi-observable. (Note that the same $\Delta G$ appears also
in the canonical decomposition.) There is no form factor measurement, which
can be used to extract $\Delta G$. 
Nevertheless, within the theoretical formulation of
DIS scatterings, $\Delta G$ and $\Delta \Sigma$ appear on the equal
footing \cite{AP1977}. 
Although the extraction of $\Delta G$ is far more difficult than
that of $\Delta \Sigma$, a great progress is under way and there is no
doubt that it will be determined more precisely in the near
future \cite{DSSV14}.
Once $\Delta \Sigma$ and $\Delta G$ are known, the quark and
gluon OAM terms in the mechanical decomposition can be extracted
from the relations
\begin{eqnarray}
 L^q_{mech} \ &=& \ J^q \ - \ \frac{1}{2} \,\Delta \Sigma, \\
 L^G_{mech} \ &=& \ J^G \ - \ \Delta G .
\end{eqnarray}
Even a direct extraction of $L_q$ might be possible through the known
relation
\begin{equation}
 L^q_{mech} \ = \ - \,\int \,d x \,x \,\, G_2 (x,0,0) ,
\end{equation}
where $G_2$ is one of the twist-3 GPD \cite{PPSS00,KP04,HMS04,HY12}.
We would thus conclude that the mechanical decomposition is an
experimentally accessible decomposition of the nucleon spin.

Let us now turn to the last decomposition, i.e. the canonical decomposition
or the Jaffe-Manohar decomposition.
We emphasize that the quark and gluon OAMs in this decomposition cannot
be extracted from the knowledge of $\Delta \Sigma$ and $\Delta G$
supplemented with that of $J^q$ and $J^G$, since \cite{Waka10,Waka11A}
\begin{eqnarray}
 L^q_{``can"} \ &\neq& \ J^q \ - \ \frac{1}{2} \,\Delta \Sigma, \\
 L^G_{``can"} \ &\neq& \ J^G \ - \ \Delta G .
\end{eqnarray}
Some years ago, Lorce and Pasquini pointed out that the canonical
quark OAM $L^\prime_q$ appearing in the Jaffe-Manohar decomposition
can be related to a moment of a Wigner distribution $F_{14}$ as \cite{LP11}
(see also \cite{Hatta12})
\begin{equation}
 L^q_{``can"} \ = \ - \,\int \,d x \,\int \,d^2 k_\perp \,
 \frac{\bm{k}^2_\perp}{M_N^2} \,\,
 F_{14} (x, \,\xi = 0, \,\bm{k}^2_\perp, \,
 \bm{k}_\perp \cdot \bm{\Delta}_\perp = 0, \,
 \bm{\Delta}^2_\perp = 0) .
\end{equation}
Soon after, however, Courtoy et al. pointed out that this Wigner function
$F_{14}$ disappears in both of the GPD (generalized parton distribution) and
TMD (transverse-momentum-dependent distribution) factorization
schemes \cite{CGHLR14}.
Since the appearance in the factorization scheme or in the cross section
formula is a necessary condition of observability or quasi-observability,
we must say that $F_{14}$ is not even a
quasi-observable, at least within our limited knowledge of DIS
measurements. One might suspect that
the canonical OAM is not observable would be connected with
the fact that it is not gauge-invariant in a strong sense.
However, the gauge-invariant definition of the gluon spin $\Delta G$
also need the idea of the physical component, while it appears in
the cross section formula within the standard collinear factorization scheme.
The underlying reason of this difference between the canonical quark OAM
and the gluon spin is still unexplained.

\section{Physical interpretation of the two OAMs of quarks}
\label{sec:5}

In the previous section, we have demonstrated that the principle-value
prescription (\ref{Eq:PV_form}) would be the most natural choice for
defining the physical component
of the gluon field. Once we accept this choice, the FSI and the ISI terms of
the average transverse momenta can be written as
\begin{equation}
 \langle k^l_\perp \rangle^{\pm LC}_{int} \ = \ \langle k^l_\perp \rangle_{pot}
 \ + \ \langle k^l_\perp \rangle_{gluon-pole} ,
\end{equation}
where
\begin{equation}
 \langle k^l_\perp \rangle_{pot} \ = \ \frac{1}{2 \,P^+} \,\langle P S_\perp \,|\,
 \bar{\psi} (0) \,\gamma^+ \,g \,A^l_{phys} (0) \,\psi (0) \,|\, P S_\perp \rangle ,
\end{equation}
corresponds to the potential momentum, while
\begin{eqnarray}
 \langle k^l_\perp \rangle_{gluon-pole} \ = \ \mp \,\frac{1}{4 \,P^+} \,
 \int_{- \,\infty}^{+ \,\infty} \,d \eta^- \,\,\langle P S_\perp \,| \,
 \bar{\psi} (0) \,\gamma^+ \,{\cal L} [0,\eta^-] \,g F^{+ l} (\eta^-) \,
 {\cal L} [\eta^-, 0] \,\psi (0) \,|\, P S_\perp \rangle ,
\end{eqnarray}
is the so-called gluon-pole term of the Efremov-Teryaev-Qui-Stermann (ETQS)
quark-gluon correlation function $\Psi_F (x, x^\prime)$ \cite{ET1985,QS1991,QS1992}, i.e.
\begin{equation}
 \langle k^l_\perp \rangle_{gluon-pole} \ = \ \frac{1}{2} \,\epsilon^{ij}_\perp \,S^j_\perp \,
 (\mp \,\pi) \,\,\int \,d x \,\Psi_F (x,x) .
\end{equation}
(We recall that $\langle k^l_\perp \rangle^{\pm LC}_{int}$can also be related to a
moment of the T-odd TMD called the Sivers function \cite{Sivers1990, Sivers1991}.)
Since it holds that
\begin{equation}
 \langle k^l_\perp \rangle_{mech} \ + \ \langle k^l_\perp \rangle_{pot}
 \ = \ \langle k^l_\perp \rangle_{``can''} ,
\end{equation}
the average transverse momenta $\langle k^l_\perp \rangle^{\pm LC}$ in
(\ref{Eq:momentum_basic}) can be expressed in either of the following two forms : 
\begin{equation}
 \langle k^l_\perp \rangle^{\pm LC} \ = \ \langle k^l_\perp \rangle_{mech}
 \ + \ \langle k^l_\perp \rangle^{\pm LC}_{int} ,
\end{equation}
or
\begin{equation}
 \langle k^l_\perp \rangle^{\pm LC} \ = \ \langle k^l_\perp \rangle_{can}
 \ + \ \langle k^l_\perp \rangle^{\pm LC}_{gluon-pole} .
\end{equation}
There is no inconsistency between these two expressions, since the PT symmetry
dictates that
\begin{equation}
 \langle k^l_\perp \rangle_{mech} \ = \ \langle k^l_\perp \rangle_{can} \ = \ 0 ,
\end{equation}
and that
\begin{equation}
 \langle k^l_\perp \rangle^{\pm LC}_{int} \ = \ \langle k^l_\perp \rangle_{gluon-pole} .
\end{equation}
We stress that $\langle k^l_\perp \rangle^{\pm LC}$ coincide with 
{\it neither} of the canonical momentum nor the mechanical one.
Such being the case, one may conclude that the idea of physical component
(or the concept of canonical momentum) plays no practically useful role in the case
of average transverse momentum.

It is therefore convenient to return to the original gauge-invariant
relation (\ref{Eq:momentum_basic}), which is independent of the idea of the
physical component of the gauge field.
For clarity, we consider below the case of the SIDIS (simi-inclusive-deep-inelastic
scattering) processes.
The relation in this case is written as
\begin{equation}
 \langle k^l_\perp \rangle^{+ \,LC} \ = \ \langle k^l_\perp \rangle_{mech}
 \ + \ \langle k^l_\perp \rangle^{+ \,LC}_{int}, \label{Eq:momentum_SIDIS}
\end{equation}
with the additional information that $\langle k^l_\perp \rangle_{mech} = 0$.
Physical interpretation of this relation should be obvious by now.
Initially, the average transverse momentum of
quarks {\it inside} the nucleon is given by 
$\langle k^l_\perp \rangle_{mech}$, which
is actually zero due to the PT symmetry. 
Through the FSI $\langle k^l_\perp \rangle^{+ \,LC}_{int}$ in the
SIDIS processes, the quark ejected by the virtual photon acquires
non-zero transverse momentum. The l.h.s. of the relation (\ref{Eq:momentum_SIDIS})
can therefore be interpreted as the transverse momentum of the quark at the
asymptotic distance, or that well outside the nucleon. 

Exactly the same interpretation must hold also for the average longitudinal OAM.
For clarity, we again confine to the case of future-pointing staple-like LC path
${\cal L} = +LC$ corresponding to the SIDIS processes.
In this case, we have the relation
\begin{equation}
 \langle L^3 \rangle^{+LC} \ = \ \langle L^3 \rangle_{mech} \ + \ 
 \langle L^3 \rangle^{+LC}_{int} . 
\end{equation}
We already know that the FSI term
$\langle L^3 \rangle^{+LC}_{int}$ coincides with the potential angular
momentum $\langle L^3 \rangle_{pot}$, so that we can also write as
\begin{equation}
 \langle L^3 \rangle^{+LC} \ = \ \langle L^3 \rangle_{mech} \ + \ 
 \langle L^3 \rangle_{pot} \ = \ \langle L^3 \rangle_{``can''} ,
\end{equation}
where the r.h.s. is the so-called gauge-invariant canonical OAM.
(Note that it is gauge-invariant only in a weak sense.)
A natural interpretation of the above relation deduced from the
average transverse momentum case is as follows.
Initially, the average OAM of quarks {\it inside} the nucleon is obviously
the manifestly gauge-invariant mechanical OAM $\langle L^3 \rangle_{mech}$,
which is generally non-zero.
Through the FSI caused by the torque of color
Lorentz force, the ejected quark acquires an additional OAM, i.e. the
potential angular momentum $\langle L^3 \rangle_{pot}$, which was
originally stored in the gluon OAM part appearing in the mechanical decomposition
of the nucleon spin. 
Consequently, the final OAM of the ejected quark is converted into the
canonical OAM.
We emphasize that this interpretation is just consistent with our previous
observation in the
Landau problem that the canonical OAM represents the total OAM, i.e. the
sum of the mechanical OAM of a particle and the OAM carried
by the electromagnetic potential. 
Now the reason why the relation $\langle L^3 \rangle^{+LC} = 
\langle L^3 \rangle_{can}$ holds should be clear.
For, according to our general rule, the average longitudinal OAM
$\langle L^3 \rangle^{+LC}$, defined by the Wigner distribution with the
gauge-link path ${\cal L} = +LC$, must represents the asymptotic
OAM of the ejected quark after leaving the spectator in the SIDIS
processes. It is only natural that this OAM of a quark well separated
from the original nucleon center reduces to the seemingly free canonical OAM,
since there is no background of color electromagnetic field in this
asymptotic distance.
It is also clear that this canonical OAM is not an
intrinsic OAM carried by the quarks inside the nucleon. Stated differently,
the canonical OAM of the Jaffe-Manohar decomposition is not an
intrinsic (or static) property of the nucleon.

Because our conclusion is fairly different from the naive picture
believed by quite a few researches in the DIS physics community, some
more explanation would be mandatory. After all, what makes our
problem delicate and complicated is the FSI or ISI,
which comes into the game through the transverse gauge-link.
This can be easily understood if one inspects the average longitudinal momentum
defined through the Wigner distribution : 
\begin{equation}
 \langle x \rangle^{\cal L} \ = \ \int \,d x \,\int \,d^2 \bm{b}_\perp \,
 \int \,d^2 \bm{k}_\perp \,\,x \,\,\rho^{\cal L} (x, \bm{b}_\perp, \bm{k}_\perp) .
\end{equation}
In this case, the integration over $\bm{b}_\perp$ and $\bm{k}_\perp$ is trivial
(the contribution from the transverse gauge-link vanishes),
and the gauge-link path dependence essentially disappears, thereby being led
to the familiar result,
\begin{equation}
 \langle x \rangle \ = \ \frac{1}{2 \,P^+} \,\langle P S \,|\,\bar{\psi} (0) \,
 \gamma^+ \,\frac{1}{i} \,D^+ (0) \,\psi (0) \,|\, P S \rangle \ = \ 
 \langle x \rangle_{mech}.
\end{equation}
This is nothing but the manifestly gauge-invariant mechanical quark
momentum $\langle x \rangle_{mech}$.
At first glance, it appears to contradict our general rule that the average
longitudinal momentum of quarks defined through the Wigner distribution should
represent the asymptotic quark momentum. There is no discrepancy, however, 
since we generally get
\begin{eqnarray}
 \langle x \rangle_{mech} \ &=& \ \frac{1}{2 \,P^+} \,
 \langle P S \,|\, \bar{\psi} (0) \,\gamma^+ \,\frac{1}{i} \,D^+ \,\psi (0) \,
 |\, P S \rangle \nonumber \\
 \ &=& \ \frac{1}{2 \,P^+} \,
 \langle P S \,|\, \bar{\psi} (0) \,\gamma^+ \,\frac{1}{i} \,D^+_{pure} \,\psi (0) \,
 |\, P S \rangle \nonumber \\
 \ &-& \ \frac{1}{2 \,P^+} \,
 \langle P S \,|\, \bar{\psi} (0) \,\gamma^+ \,A^+_{phys} (0) \,\psi (0) \,
 |\, P S \rangle \nonumber \\
 \ &=& \ \langle x \rangle_{``can"} \ - \ \langle x \rangle_{pot} ,
\end{eqnarray}
and since we know that the FSI or the potential momentum term vanishes identically,
i.e. $\langle x \rangle_{pot} = 0$.
(This is manifest in the LC gauge $A^+ = A^+_{phys} = 0$, and it is true
also in general gauge \cite{Jaffe01}.)
Namely, due to the cancellation of the FSI for the collinear momentum case,
there is no difference between the canonical and mechanical momenta,
\begin{equation}
 \langle x \rangle_{mech} \ = \ \langle x \rangle_{``can"} .
\end{equation}
In this case, one is therefore allowed to say that either of the canonical or mechanical
momentum is partonic and at the same time either represents the intrinsic property of
the nucleon. 

As explained above, this is clearly not the case for the OAM of
quarks in the nucleon. What would be an underlying physical reason for this difference ?
It can be easily understood from our consideration of the cyclotron motion of
a charged particle in sect.II. A generation of non-zero
orbital angular momentum in the stationary nucleon state necessarily requires circular motion
of quarks. This circular motion cannot be a free (or translational) motion in any sense.
One might say that this is certainly true for the mechanical OAM but that the
same argument does not apply to the canonical OAM, since the latter looks like
the OAM of free quarks.
However, what meaning does it have to say that such an orbital angular
momentum well outside the nucleon is partonic ? 

After all, a natural conclusion is that neither of the canonical OAM nor the mechanical
OAM cannot be interpreted as partonic. 
Both are intrinsically twist-3 quantities.
To convince it, we recall the following relation derived by
Hatta and Yoshida \cite{HY12}.
\begin{equation}
 \Phi_D (x_1,x_2) \ = \ P \,\frac{1}{x_1 - x_2} \,\Phi_F (x_1,x_2) \ + \ 
 \delta (x_1 - x_2) \,L^q_{can} (x_1) , \label{Eq:HY}
\end{equation}
where $\Phi_D (x_1,x_2)$ is the D-type quark-gluon correlation function of twist-3
defined by
\begin{eqnarray}
 &\,& \int \frac{d \lambda}{2 \,\pi} \,\int \frac{d \mu}{2 \,\pi} \,
 e^{\,i \,\lambda \,x_1} \,e^{\,i \,\mu (x_2 - x_1)} 
 \,\,\langle p^\prime, s^\prime \,|\, \bar{\psi} (0) \,\gamma^+ \,
 {\cal L} [0,\mu] \,D^i (\mu) \,{\cal L} [\mu, \lambda] \,\psi (\lambda) \,|\, p, s \rangle
 \nonumber \\
 &\,& \hspace{4mm} \ = \ \epsilon^{+ i \rho \sigma} \,S_\rho \,\Delta_{\perp, \sigma} \,
 \Phi_D (x_1,x_2) \ + \ \cdots , 
\end{eqnarray}
while $\Phi_F (x_1,x_2)$ is the  $F$-type quark-gluon correlation functions defined by
\begin{eqnarray}
 &\,& \int \frac{d \lambda}{2 \,\pi} \,\int \frac{d \mu}{2 \,\pi} \,
 e^{\,i \,\lambda \,x_1} \,e^{\,i \,\mu (x_2 - x_1)} 
 \,\,\langle p^\prime, s^\prime \,|\, \bar{\psi} (0) \,\gamma^+ \,
 {\cal L} [0,\mu] \,g \,F^{+ i} (\mu) \,{\cal L} [\mu, \lambda] \,\psi (\lambda) \,|\, p, s \rangle
 \nonumber \\
 &\,& \hspace{4mm} \ = \ P^+ \,\epsilon^{ij}_\perp \,S^j_\perp \,\Psi_F (x_1, x_2) \ + \ 
 \epsilon^{ij}_\perp \,\Delta^j_\perp \,\,S^+ \,\Phi_F (x_1,x_2) \ + \ \cdots . 
\end{eqnarray}
(Note that $\Psi_F (x_1, x_2)$ here is more familiar 
Efremov-Teryaev-Qui-Sterman (ETQS)
function, while another correlation function $\Phi_F (x_1,x_2)$ appears in (\ref{Eq:HY}).) 
$L^q_{can} (x_1)$ in (\ref{Eq:HY}) is the canonical OAM density given
as \cite{HY12}
\begin{eqnarray}
 L_{can} (x) \ &=& \ x \,\int_x^{\epsilon (x)} \,\frac{d x^\prime}{x^\prime} \,
 ( H_q (x^\prime) \ + \ E_q (x^\prime) ) \ - \ s \,
 \int_x^{\epsilon (x)} \,\frac{d x^\prime}{x^{\prime 2}} \,\tilde{H}_q (x^\prime) \nonumber \\
 \ &-& \ x \,\int_x^{\epsilon (x)} \,d x_1 \,\int_{-1}^1 \, d x_2 \,\,\Phi_F (x_1, x_2) \,
 \frac{3 \,x_1 - x_2}{x_1^2 \,(x_1 - x_2)^2} \nonumber \\
 \ &-& \ x \,\int_x^{\epsilon (x)} \,d x_1 \,\int_{-1}^1 \,d x_2 \,\,\tilde{\Phi}_F (x_1,x_2) \,
 P \,\frac{1}{x_1^2 \,(x_1 - x_2)} ,
\end{eqnarray}
with $\tilde{\Phi}_F (x_1,x_2)$ being a $F$-type quark-gluon correlation function defined
by
\begin{eqnarray}
 &\,& \int \frac{d \lambda}{2 \,\pi} \,\int \frac{d \mu}{2 \,\pi} \,
 e^{\,i \,\lambda \,x_1} \,e^{\,i \,\mu (x_2 - x_1)} 
 \,\, \langle p^\prime, s^\prime \,|\, \bar{\psi} (0) \,\gamma^+ \,\gamma_5 \,
 {\cal L} [0,\mu] \,g \,F^{+ i} (\mu) \,{\cal L} [\mu, \lambda] \,\psi (\lambda) \,|\, p, s \rangle
 \nonumber \\
 &\,& \hspace{4mm} \ = \ P^+ \,\epsilon^{ij}_\perp \,S^j_\perp \,\tilde{\Psi}_F (x_1, x_2) \ + \ 
 \epsilon^{ij}_\perp \,\Delta^j_\perp \,\,S^+ \,\tilde{\Phi}_F (x_1,x_2) \ + \ \cdots . 
\end{eqnarray}
Our interest here is only the integrated OAMs, since we think that
the density level
decomposition needs more satisfactory understanding of the role of
surface terms, 
which we do not believe has been cleared up yet. Then, using the relations
\begin{eqnarray}
 L^q_{mech} \ &=& \ \int \,d x_1 \,\int \,d x_2 \,\Phi_D (x_1,x_2) , \\
 \ L_{pot} \ &=& \ \int \,d x_1 \,\int \,d x_2 \,\,P \,\frac{1}{x_1 - x_2} \,\Phi_F (x_1, x_2) , 
\end{eqnarray}
as well as the symmetries of the correlation functions $\Phi_F (x_1,x_2)$ and
$\tilde{\Phi}_F (x_1,x_2)$, 
\begin{eqnarray}
 \Phi (x_1, x_2) \ = \ \Phi (x_2, x_1), \ \ \ \ \tilde{\Phi} (x_1, x_2) \ = \ - \,\tilde{\Phi} (x_2, x_1),
\end{eqnarray}
one readily obtains
\begin{eqnarray}
 L_{can} \ &=& \ \int \,d x_1 \,\int \,d x_2 \,\,\delta (x_1 - x_2) \,L_{can} (x_1) \nonumber \\
 \ &=& \ \frac{1}{2} \,\int \,d x \,\,x \,(H_q (x) + E_q (x)) \ - \ 
 \frac{1}{2} \,\int \,d x \,\tilde{H}_q (x) \ + \ L_{pot} .
\end{eqnarray}
The fact that the potential OAM $L_{pot}$ is related to the genuine twist-3 quark-gluon
$\Phi_F (x_1, x_2)$ correlation function is nothing surprising, since we already explained our
interpretation that $L_{pot}$ is just the FSI in the SIDIS processes or ISI in the Drell-Yan
processes. Noteworthy fact here is that the genuine twist-3 piece of $L_{can}$ is
precisely canceled by that of $L_{pot}$ in the combination $L_{mech} = L_{can} - L_{pot}$.
This result could be anticipated from the famous Ji sum rule \cite{Ji1998},
which is given only with the twist-2 quantities as
\begin{equation}
 L_{mech} \ = \ \frac{1}{2} \,\int \,d x \,\,x \,(H_q (x) \ + \ E_q (x)) \ - \ 
 \frac{1}{2} \,\int \,d x \,\tilde{H}_q (x) .
\end{equation}
Still interesting is the fact that this cancellation reminds us of the
observation in the Landau problem
that the quantum number $m$ dependence of the canonical OAM and
that of the potential angular momentum are just canceled and the mechanical
OAM is independent of this unphysical quantum number $m$, 
the eigenvalue of the canonical OAM operator.

\section{Concluding remarks}
\label{sec:6}

The main objective of the present paper is to get a clear understanding on
the physical meaning of the two existing decompositions of the nucleon spin, 
i.e. the canonical and mechanical decompositions.
Needless to say, when one talks about the decomposition of the nucleon spin, one is
tacitly supposing in mind the intrinsic spin structure of the nucleon.
As we have shown, what meets this requirement is the mechanical decomposition
not the canonical decomposition also known as Jaffe-Manohar decomposition.
In fact, the canonical quark OAM represents the OAM of an
ejected quark in the SIDIS processes. Putting in other words, it stands for the
OAM of a quark well {\it outside} the nucleon. How can one think of it as representing
an intrinsic (static) structure of the nucleon ?

There is wide-spread misbelief in the DIS physics community
that the canonical OAM just matches the partonic picture of quark motion in the nucleon.
This misunderstanding partially comes from the fact that, for the collinear quark
(or gluon) momentum fraction, there is no difference between the canonical and
mechanical momenta due to the cancellation of the final state interaction
in the inclusive DIS processes. 
In this case, one can say that either of the canonical or mechanical momentum
is partonic and besides either represents the intrinsic property of the nucleon.
This is not the case for the orbital angular momentum of quarks, however. 
The reason for it is clear by now from our present analysis.
The generation of non-zero orbital angular momentum inside the stationary nucleon
state necessarily requires circular motion of a particle. 
The point is that this circular motion cannot be a free motion in any sense.
In fact, we showed that neither of the canonical OAM nor the mechanical OAM can be
partonic. They are intrinsically twist-3 objects. Still, one should pay close attention to
the vital difference between these two OAMs. An obvious superiority of the mechanical OAMs
are that they are observables (or at least a quasi-observables) within the framework
of the GPD factorization scheme.  On the other hand, the $F_{14}$ sum rule,
which was once believed to provide us with a hope to experimentally access the
canonical OAM of quarks, is questioned now since the Wigner distribution $F_{14}$
does not appear in either of the GPD (collinear) and TMD factorization schemes.
In that sense, one might be able to say that it is not even a quasi-observable,
at least according to our present knowledge of the method of measurement
based on the pQCD framework. 
In our opinion, this proves the validity of our claim of long years, which
advocates the superiority of the mechanical type decomposition of the
nucleon spin over the canonical one either from the physical viewpoint or
from the observational viewpoint.

\vspace{3mm}
\begin{acknowledgments}
The author greatly acknowledges enlightening discussion with C.~Lorc\'{e} 
and S.C.~Tiwari. He also appreciate many helpful discussions with K. Tanaka
and S.~Kumano, especially on the definition of the longitudinally polarized
gluon distributions.
\end{acknowledgments}





\end{document}